\documentclass[11pt,a4paper]{article}
\usepackage{jcappub}

\usepackage{amssymb,amsmath,amstext,amsfonts}
\usepackage{bbold}
\usepackage{bm}
\usepackage{graphicx}
\usepackage{xcolor}
\usepackage[normalem]{ulem}
\usepackage{overpic}
\usepackage{subcaption}
\usepackage{colortbl}
\usepackage{hyperref, cleveref}
\usepackage{soul}

\newcommand{\beq}{\begin{equation}}
\newcommand{\eeq}{\end{equation}}
\newcommand{\bal}{\begin{aligned}}
\newcommand{\eal}{\end{aligned}}

\newcommand{\Pzeta}{\mathcal{P}_{\zeta}}

\newcommand{\Pbar}{\overline{\mathcal{P}}_{\zeta}}

\newcommand{\omegalin}{\omega_{\textrm{lin}}}
\newcommand{\omegagwlin}{\omega_{\textrm{lin}}^{\textsc{gw}}}
\newcommand{\omegalog}{\omega_{\textrm{log}}}

\newcommand{\Alog}{A_{\textrm{log}}}
\newcommand{\philog}{\phi_{\textrm{log}}}

\newcommand{\OGW}{\Omega_\textrm{GW}}
\newcommand{\ObarGW}{\overline{\Omega}_\textrm{GW}}

\newcommand{\fref}{f_\textrm{ref}}
\newcommand{\etap}{\gamma}
\newcommand{\C}{\mathcal{C}}

\title{Detecting primordial features with LISA}

\author[a,b,c]{Jacopo Fumagalli,}
\author[d]{Mauro Pieroni,}
\author[a]{S\'{e}bastien Renaux-Petel,}
\author[a,1]{Lukas~T. Witkowski}
\note{Corresponding author}

\affiliation[a]{Institut d'Astrophysique de Paris, GReCO, UMR 7095 du CNRS et de Sorbonne Universit\'{e}, 98bis
boulevard Arago, 75014 Paris, France}
\affiliation[b]{Instituto de F\'{i}sica Te\'{o}rica UAM/CSIC, Calle Nicol\'as Cabrera 13-15, Cantoblanco E-28049 Madrid, Spain}
\affiliation[c]{
Departamento de F\'{i}sica Te\'{o}rica, Universidad Aut\'{o}noma de Madrid (UAM), Campus de Cantoblanco, E-28049 Madrid, Spain
}
\affiliation[d]{Theoretical Physics, Blackett Laboratory, Imperial College, London, SW7 2AZ, United Kingdom}

\emailAdd{jacopo.fumagalli@uam.es}
\emailAdd{m.pieroni@imperial.ac.uk}
\emailAdd{renaux@iap.fr}
\emailAdd{lukas.witkowski@iap.fr}

\abstract{Oscillations in the frequency profile of the stochastic gravitational wave background are a characteristic prediction of small-scale features during inflation. In this paper we present a first investigation of the detection prospects of such oscillations with the upcoming space-based gravitational wave observatory LISA. As a proof of principle, we show for a selection of feature signals that the oscillations can be reconstructed with LISA, employing a method based on principal component analysis. We then perform a Fisher forecast for the parameters describing the oscillatory signal. For a sharp feature we distinguish between the contributions to the stochastic gravitational wave background induced during inflation and in the post-inflationary period, which peak at different frequencies. We find that for the latter case the amplitude of the oscillation is expected to be measurable with $< 10\%$ accuracy if the corresponding peak satisfies $h^2 \Omega_\textrm{GW} \gtrsim 10^{-12}$-$10^{-11}$, while for inflationary-era gravitational waves a detection of the oscillations requires a higher peak amplitude of $h^2 \Omega_\textrm{GW}$, as the oscillations only appear on the UV tail of the spectrum. For a resonant feature the detection prospects with LISA are maximised if the frequency of the oscillation falls into the range $\omegalog = 4$ to $10$. Our results confirm that oscillations in the frequency profile of the stochastic gravitational wave background are a worthwhile target for future detection efforts and offer a key for experimentally testing inflation at small scales.}

\begin{document}
\hfill{\flushright {IFT-UAM/CSIC-21-143}}
\maketitle

\section{Introduction}
Gravitational Waves (GWs), once produced, propagate effectively freely through the Universe owing to the weakness of the gravitational coupling. As a result, GWs from the early Universe can contain information about this period that is effectively untainted by the subsequent cosmological history. With operational ground-based detectors and space-based experiments a decade away, physics is acquiring a new `sense' for testing our understanding of cosmology \cite{Caprini:2018mtu} and astrophysics \cite{Regimbau:2011rp}. 

What makes GW astronomy particularly interesting is that it can grant experimental access to processes and cosmological eras that so far have remained dark. One such era is the later stage of cosmic inflation, after the modes accessible to Cosmic Microwave Background (CMB) and Large Scale Structure (LSS) experiments have exited the Hubble radius. For these small scales ($\ll 1 \textrm{ Mpc}$) the CMB and LSS constraints on scalar fluctuations do not apply, which for scales $\gtrsim 1 \textrm{Mpc}$ demand that the fluctuations are effectively Gaussian and have a nearly scale invariant power spectrum with amplitude $\Pzeta \sim 10^{-9}$. As a result, at this later stage inflation can depart significantly from the canonical single-field slow-roll paradigm. 

What is interesting for GW observatories is that these departures from single-field slow-roll can lead to an enhancement of scalar fluctuations by orders of magnitude compared to their amplitude at CMB scales. As scalar fluctuations couple to tensor modes at second order in perturbations, they induce gravitational waves during inflation and again after inflation as they reenter the horizon (see \cite{Domenech:2021ztg} for a review focusing on the latter case), both of which contribute to the Stochastic GW Background (SGWB). For these scalar-induced GWs the energy density fraction today scales as $\OGW \sim 10^{-5} \Pzeta^2$, so that an enhancement in $\Pzeta$ is crucial for detectability. Besides GWs, a complementary observational signal for inflation models with enhanced scalar fluctuations is the abundance of primordial black holes, which are produced as the fluctuation-induced overdensities collapse at horizon reentry (see \cite{Sasaki:2018dmp,Carr:2020gox,Carr:2020xqk,Green:2020jor} for recent reviews).

Inflation models that lead to an enhancement of scalar fluctuations at small scales frequently also exhibit another characteristic property in the form of oscillations in the scalar power spectrum, also known as \emph{features} (see  e.g.~\cite{Slosar:2019gvt} for a review). In particular, one can identify \emph{sharp features}, caused by a sudden transition in a background quantity during inflation and characterised by oscillations linear in the frequency $f$, and \emph{resonant features}, associated with an oscillation in a background quantity and exhibiting a periodic modulation in $\ln(f)$. What is interesting for GW astronomy is that, as has been seen in \cite{Fumagalli:2020nvq, Braglia:2020taf, Fumagalli:2021cel, Dalianis:2021iig, Witkowski:2021raz, Fumagalli:2021mpc, Iacconi:2021ltm}, the scalar-induced GW spectrum related to these scales also exhibits oscillations in the frequency profile. For example, the post-inflationary era induced contribution to the energy density fraction in GWs has been shown to take the form \cite{Fumagalli:2020nvq, Fumagalli:2021cel,Witkowski:2021raz}:
\begin{align}
    \label{eq:GW-sharp-template} 
    \textrm{Sharp:} \quad & h^2 \Omega_{\textrm{GW}}(f) = h^2 \overline{\Omega}_{\textrm{GW}}(f) \Big[1+ \mathcal{A}_\textrm{lin} \cos \big(\omegagwlin f + \theta_\textrm{lin}  \big) \Big] \, , \\
    \label{eq:resonant-template-intro}
    \textrm{Resonant:} \quad & h^2 \Omega_{\textrm{GW}}(f) = h^2 \overline{\Omega}_{\textrm{GW}}(f) \Big[1+ \mathcal{A}_{\textrm{log},1} \cos \big(\omegalog \ln (f/f_\textrm{ref}) + \theta_{\textrm{log},1} \big) \\
    \nonumber & \hphantom{h^2 \Omega_{\textrm{GW}}(f) = h^2 \overline{\Omega}_{\textrm{GW}}(f) \Big[1} + \mathcal{A}_{\textrm{log},2} \cos \big(2 \omegalog \ln (f/f_\textrm{ref}) + \theta_{\textrm{log},2} \big) \Big] \, ,
\end{align}
with $\ObarGW(f)$ a model-dependent but smooth envelope. The inflationary-era induced contribution due to a sharp feature, also exhibits an oscillatory frequency profile \cite{Fumagalli:2021mpc}, with the corresponding template given in \eqref{eq:GW-inf-template}. This offers the exciting prospect that small-scale features during inflation can be tested through their characteristic oscillatory contribution to the SGWB. 

While oscillations in the frequency profile of the SGWB are thus a worthwhile target for future detection efforts, their detection prospects are currently unknown. What makes a measurement of the oscillations potentially challenging is that the oscillation amplitude in \eqref{eq:GW-sharp-template} and \eqref{eq:resonant-template-intro} is typically less than unity, even if the corresponding scalar power spectrum exhibits $\mathcal{O}(1)$ oscillations. For a sharp feature the oscillation amplitude is typically $\mathcal{A}_\textrm{lin} \lesssim \mathcal{O}(10 \%)$ and at most $\lesssim 40 \%$ \cite{Fumagalli:2020nvq, Witkowski:2021raz}. For a resonant feature the oscillation amplitudes $\mathcal{A}_{\textrm{log},1,2}$ can be $\mathcal{O}(1)$ for small values of $\omegalog$, but then quickly drop to $\mathcal{A}_{\textrm{log},1,2} \lesssim \mathcal{O}(10 \%)$ for larger values of $\omegalog$ \cite{Fumagalli:2021cel}. In contrast, inflationary-era GWs due to a sharp feature exhibit $\mathcal{O}(1)$ oscillations in the frequency profile, but the oscillations only occur on the UV tail of the spectrum \cite{Fumagalli:2021mpc}, making detection again more difficult.

The goal of this work is thus to perform a first investigation into the detectability of oscillations in the SGWB with the Laser Interferometer Space Antenna (LISA) \cite{LISA:2017pwj, LISA:web}. The approved experimental make-up of LISA is a constellation of three satellites in a nearly equilateral triangle with an arm length of 2.5 million km and most sensitive to GWs in the mHz frequency range. This frequency range implies that LISA will be sensitive to features during inflation that occur $\sim 30$ $e$-folds after the exit of CMB modes, see e.g.~\cite{Fumagalli:2020nvq}. 
First, we will apply the reconstruction technique based on a Principal Component Analysis (PCA) from \cite{Pieroni:2020rob} to mock data to show that the oscillations in $h^2 \OGW(f)$ can in principle be reconstructed for reasonable parameter choices. Given this proof of principle regarding detection, we then perform a forecast based on the Fisher Information Matrix (FIM), focussing on the parameters characterising the oscillation. Using the FIM analysis to assess the predicted accuracy, we then perform a scan over the parameter space given by the oscillation variables and the overall amplitude of $h^2\OGW$. The result are maps of this parameter space displaying the $\%$-accuracy in the quantity of interest as well as the Signal-to-Noise Ratio (SNR). 

In this first investigation we do not consider astrophysical foregrounds, and we also take the apparatus noise as fixed and do not include the noise parameters in the FIM analysis. Another limitation of the FIM-based approach is that it does not give information about the best-fit theoretical model for a given dataset. The goal of this work is thus not to provide a conclusive analysis of the detection prospects for oscillations due to features. Instead, in this work we aim to provide (\emph{i}) a proof of principle that oscillations due to small-scale features during inflation could be detected by LISA, and, assuming a template for the oscillatory features, (\emph{ii}) stake out the most promising regions of parameter space where such a detection may be possible. For this promising regions we then envision future work using more sophisticated methods like the binning technique of \cite{Caprini:2019pxz, Flauger:2020qyi} and a proper MCMC sampling of the parameter space, to conclusively determine the detectability.  

This paper is organised as follows. In sec.~\ref{sec:templates} we collect the relevant templates for $h^2 \OGW(f)$ for scalar-induced contributions to the SGWB due to small-scale features during inflation. In sec.~\ref{sec:PCA} we briefly review the PCA method of \cite{Pieroni:2020rob} and apply it to a selection of example signals. In sec.~\ref{sec:Fisher} we briefly introduce the concept of a Fisher forecast, and then apply it in turn to the templates from sec.~\ref{sec:templates}. Sec.~\ref{sec:conclusions} contains the conclusion and in appendix \ref{app:resonant-template-coeffs} we collect expressions used in the definition of the templates.

\section{Primordial features in the stochastic gravitational wave background: templates}
\label{sec:templates}
In this section we collect the templates for $h^2 \OGW(f)$ due to primordial features during inflation, state the underlying assumptions and briefly summarise how the parameters describing the templates are related to the parameters that characterise the primordial feature. For further details and the derivation of these expression we refer to \cite{Fumagalli:2020nvq,Fumagalli:2021cel, Witkowski:2021raz, Fumagalli:2021mpc} where these templates were first derived. 

Primordial features induce GWs both during inflation and again in the post-inflationary era, when the relevant scalar modes re-enter the horizon. The total GW spectrum $h^2 \OGW(f)$ is the sum of the contributions from these two phases together with a cross-term that mixes the two. Here we will consider the situation where either one of the two production phases gives the dominant contribution to $h^2 \OGW(f)$, so that the contribution from the other and the mixed term can be ignored. We further differentiate between sharp and resonant features, characterised by the presence of oscillations in the scalar power spectrum taking the form:\footnote{As the argument we already use the frequency $f$ more prevalent in the study of GWs rather than the comoving wavenumber $k$ more common in an inflationary context. In any case, the relation between the two is linear.}
\begin{align}
    \label{eq:P-sharp-intro}
    \textrm{Sharp:} \quad & \Pzeta(f)=\Pbar(f) \Big[1 + A_{\textrm{lin}} \cos \big(\omega_{\textrm{lin}} f + \phi_{\textrm{lin}} \big) \Big] \, , \\
    \label{eq:P-resonant-intro}
    \textrm{Resonant:} \quad & \Pzeta(f)=\Pbar(f) \Big[1 + \Alog \cos \big(\omegalog \ln(f/f_\textrm{ref}) + \philog \big) \Big] \, ,
\end{align}
where $\Pbar(f)$ is a possibly peaked, but otherwise smooth envelope, and $\fref$ is an arbitrary reference frequency. The post-inflationary contribution to the SGWB was investigated for sharp features in \cite{Fumagalli:2020nvq,Witkowski:2021raz} and for resonant features in \cite{Fumagalli:2020nvq,Fumagalli:2021cel,Witkowski:2021raz}, while the inflationary-era contribution was recently studied for sharp features in \cite{Fumagalli:2021mpc}. Note that oscillations of sharp and resonant feature type can also occur together, e.g.~when a heavy field starts oscillating after a sudden initial perturbation during inflation, which is referred to as a primordial standard clock feature, see e.g.~\cite{Chen:2011zf,Chen:2012ja,Chen:2014joa,Chen:2014cwa}. The corresponding post-inflationary era contribution to the SGWB in models exhibiting this feature was considered in \cite{Braglia:2020taf}.   

\begin{figure}[t]
\centering
\begin{overpic}[width=0.7\textwidth]{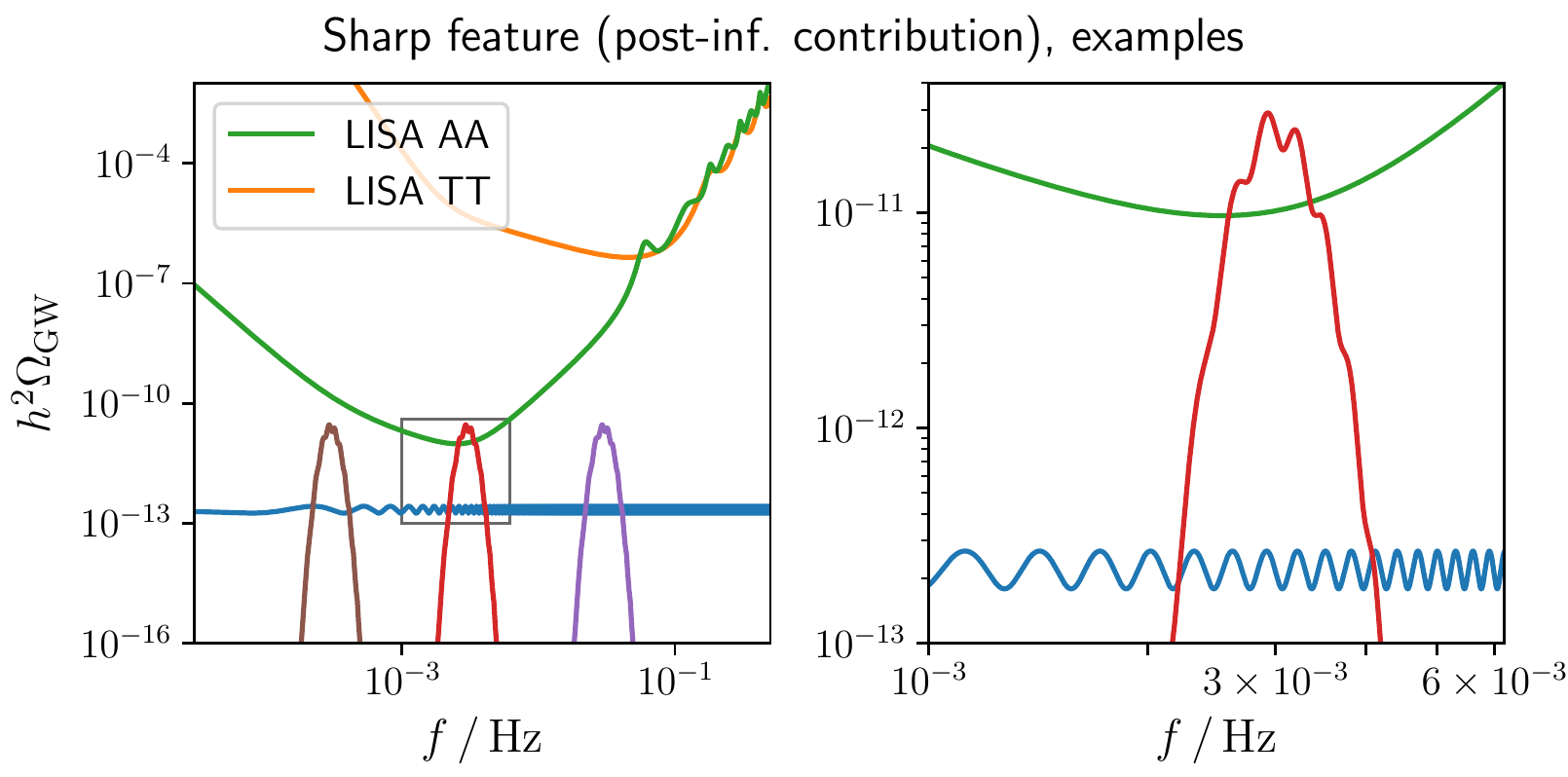}
\end{overpic}
\caption{Examples of $h^2 \OGW(f)$ for the post-inflationary contribution to the SGWB due to a sharp feature, given by the template \eqref{eq:GW-sharp-template}, together with the LISA AA and TT noise curves. The blue curve is for a flat envelope and the three peaked curves are for a lognormal envelope as given in \eqref{eq:lognorm-env}. All examples have an oscillation amplitude $\mathcal{A}_\textrm{lin}= 20 \%$. For the peaked examples the model parameters satisfy $\omegagwlin f_\star \Delta \approx 2 \pi$. To better display the modulations on the peaks, the right panel zooms in on the boxed area in the left panel.}
\label{fig:sharp_post_examples}
\end{figure}

\subsection{Sharp feature: post-inflationary contribution}
\label{sec:templates-sharp-post}
The post-inflationary contribution to $h^2 \OGW(f)$ due to a sharp feature is well-approximated by the template \eqref{eq:GW-sharp-template}, i.e.~the GW spectrum exhibits sinusoidal oscillations in $f$ modulating an otherwise smooth envelope $h^2 \ObarGW(f)$. For a sharp feature this envelope is typically peaked, and the template \eqref{eq:GW-sharp-template} is valid for frequencies in the vicinity of this peak.\footnote{For example, the template \eqref{eq:GW-sharp-template} does not correctly capture the IR tail, which does not exhibit modulations.} In this frequency range the amplitude $\mathcal{A}_\textrm{lin}$ can be taken as constant, which is what we do here. The precise shape of the envelope $h^2 \ObarGW(f)$ is model-dependent and will differ between different realisations of a sharp feature. However, when close to the detection threshold, which is the principal regime of interest in this work, we expect the precise shape of this peak to be less important, as only the frequencies close to the global maximum will be ``visible'' to the experiment. As a representative example for a peaked envelope here we choose a lognormal peak. In addition, to disentangle the detectability of the oscillations in \eqref{eq:GW-sharp-template} from properties regarding the shape of the envelope we will also consider the toy example with $h^2 \ObarGW =\textrm{const}$. Thus, in this work we will consider the following two choices for the envelope in \eqref{eq:GW-sharp-template}:  
\begin{align}
\label{eq:flat-env}    \textrm{Flat:} \qquad & h^2 \ObarGW(f) = h^2 \ObarGW = \textrm{const.} \, , \\
\label{eq:lognorm-env}    \textrm{Lognormal:} \qquad & h^2 \ObarGW(f) =  h^2 \ObarGW \, \exp \bigg(- \frac{1}{2 \Delta^2} \bigg(\ln \frac{f}{f_\star} \bigg)^2 \bigg) \, ,
\end{align}
where $f_\star$ denotes the maximum of the envelope and $\Delta$ is a dimensionless ``standard deviation''. Putting everything together, the model parameters are $(h^2 \ObarGW, \, \mathcal{A}_\textrm{lin}, \, \omegagwlin, \, \theta_\textrm{lin}, \,  f_\star, \, \Delta)$, with the latter two absent for the toy example with a flat envelope. In fig.~\ref{fig:sharp_post_examples} we show examples of the sharp feature template \eqref{eq:GW-sharp-template} for both a flat and lognormal envelope.

The value of $\omegagwlin$ is related to the frequency $\omegalin$ in the scalar power spectrum \eqref{eq:P-sharp-intro} as $\omegagwlin = c_s^{-1} \omegalin$, see \cite{Witkowski:2021raz}, with $c_s$ the propagation speed of density fluctuations, which we take as constant during the production of GWs.\footnote{The standard case of GWs induced during a period of radiation domination corresponds to $w=c_s^2=1/3$.} Hence, there is a degeneracy between $\omegalin$ and $c_s$ for a given value of $\omegagwlin$. The amplitude of the oscillations in \eqref{eq:GW-sharp-template} typically cannot be much larger than $\mathcal{A}_\textrm{lin} \sim \mathcal{O}(10 \%)$ due to the non-linearities in processing scalar into tensor fluctuations \cite{Fumagalli:2020nvq}. The largest value of $\mathcal{A}_\textrm{lin}$ occurs when the GW spectrum only exhibits a few periods of oscillation across the peak. For the envelope choice in \eqref{eq:lognorm-env} this corresponds to $\omegagwlin f_\star \Delta \sim \mathcal{O}(1) \cdot 2 \pi$.\footnote{Increasing the value of the combination $\omegagwlin f_\star \Delta$ the amplitude $\mathcal{A}_\textrm{lin}$ decreases.} In this case the amplitude can be as large as $\mathcal{A}_\textrm{lin} \approx 20 \%$ for GWs induced during a period of radiation domination, i.e.~$w=c_s^2=1/3$, see \cite{Fumagalli:2020nvq,Witkowski:2021raz}. Even larger values are possible for a stiffer equation of state $w=c_s^2>1/3$, with $\mathcal{A}_\textrm{lin} \approx 35 \%$ observed for $w=c_s^2=9/10$, see \cite{Witkowski:2021raz}.\footnote{This fact can be used to break the degeneracy between $c_s$ and $\omegalin$. E.g.~if an amplitude $\mathcal{A}_\textrm{lin} > 20 \%$ is observed, this would imply that $c_s^2 > 1/3$.} This shows that there are cross-correlations between the parameters describing the template, e.g.~between the combination $\omegagwlin f_\star \Delta$ and the amplitude $\mathcal{A}_\textrm{lin}$. For simplicity, in the analysis presented in this work we will not make use of these cross-correlations when re-constructing the template. However, when proposing benchmark models we pick parameters that are on the whole mutually consistent. The exception is the flat envelope case, which will still be instructive as the most optimistic scenario for a detection of oscillations due to a sharp feature.

Also note that the combination $\omegagwlin f_\star \Delta$ cannot be much smaller than $2 \pi$, as this would imply that one period of the oscillation is wider than the extent of the peak in the envelope, violating the assumption behind the template \eqref{eq:GW-sharp-template} of an envelope modulated by oscillations.

\begin{figure}[t]
\centering
\begin{overpic}[width=0.7\textwidth]{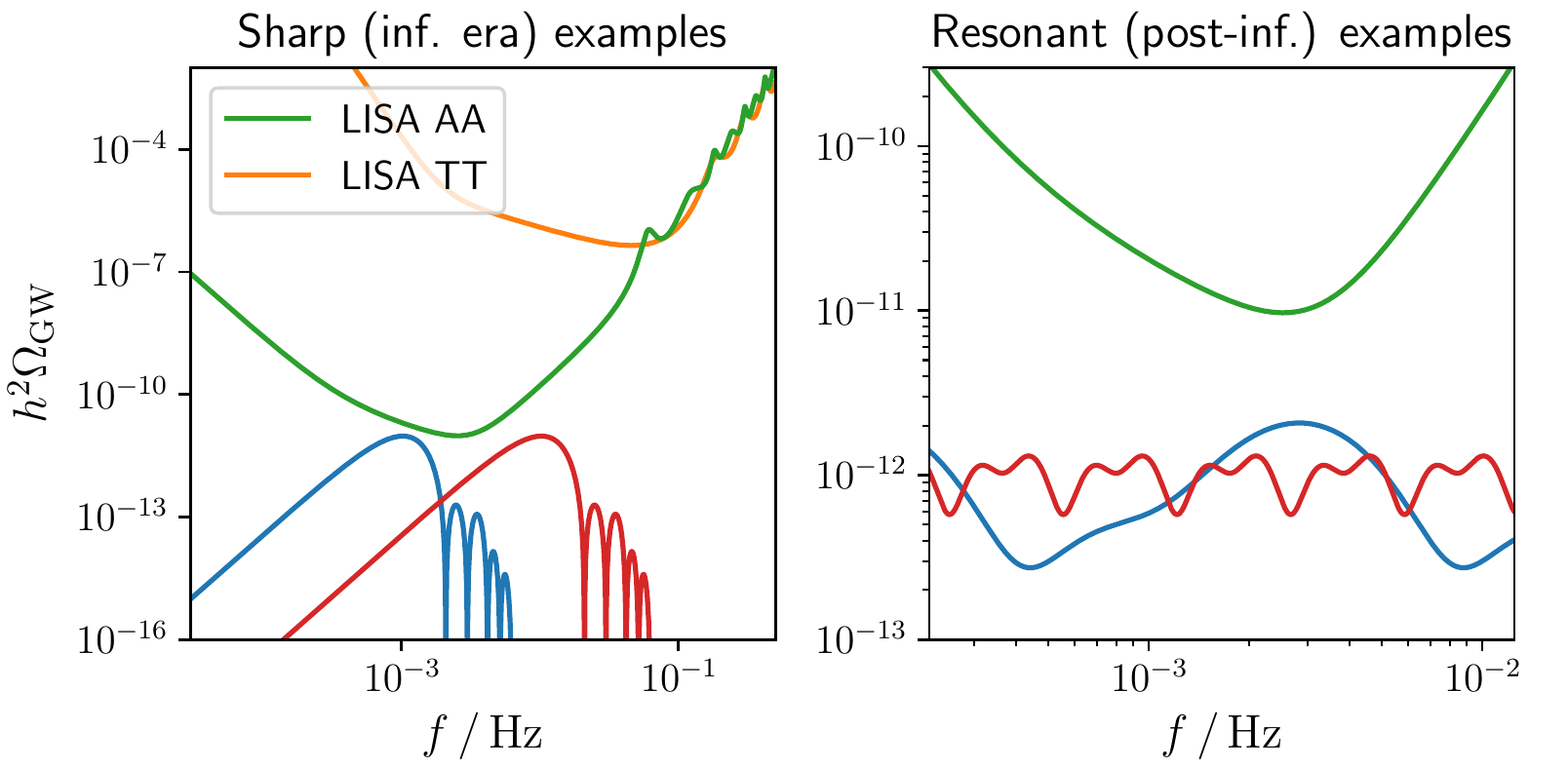}
\end{overpic}
\caption{Examples of $h^2 \OGW(f)$ for the inflationary era contribution to the SGWB due to a sharp feature (left panel) and the post-inflationary era contribution due to a resonant feature (right panel), together with the LISA AA and TT noise curves. The corresponding templates are given in \eqref{eq:GW-inf-template} and \eqref{eq:resonant-template-A}, respectively. The model parameters for the left panel plots are given in \eqref{eq:sharp-inf-benchmark-I} for the blue curve and in \eqref{eq:sharp-inf-benchmark-II} for the red curve. The model parameters for the right panel plots are $h^2 \ObarGW=10^{-12}$, $A_\textrm{log}=1$, $\phi_\textrm{log}=0$ and $\omegalog=2.1$ for the blue curve or $\omegalog=8$ for the red curve.}
\label{fig:inf_sharp_res_post_examples}
\end{figure}

\subsection{Sharp feature: inflation-era contribution}
\label{sec:templates-sharp-inf}
The spectrum of GWs induced during inflation by a sharp feature has been analysed in \cite{Fumagalli:2021mpc}. As long as the sharp feature enhances a narrow band of modes, synonymous with a narrow peak in $\Pzeta$, the corresponding contribution to $h^2 \OGW$ can be written as:
\begin{align}
\nonumber   h^2 \OGW(f) = \bigg(\frac{h^2 \ObarGW}{0.0065} \bigg) & \frac{1}{(\omegalin f)^3} \bigg(1 - \frac{(\omegalin f)^2}{16 \etap^2} \bigg)^2  \times \\
\label{eq:GW-inf-template}   & \times \Bigg[\sin \big(\tfrac{1}{2} \omegalin f \big) \,  - \,  \frac{4  \big( 1 - \cos \big( \tfrac{1}{2} \omegalin f \big) \big)}{\omegalin f}  \Bigg]^2 \Theta \big(4 \etap/\omegalin - f \big) \, ,
\end{align}
with $\etap \gg 1$. Two examples of this template are shown in the left panel of fig.~\ref{fig:inf_sharp_res_post_examples} (see the caption for the parameters used). In \eqref{eq:GW-inf-template} we have chosen $h^2 \ObarGW$ to denote the value at the maximum which is located at $f_\textrm{max} \simeq 6.223/\omegalin$. The parameter $\etap$ controls the `width' in $f$ of the envelope of the spectrum and the separation between $f_\textrm{max}$ and the maximum $f_\star$ of the envelope of the post-inflationary GW spectrum due to the same underlying sharp feature, with $f_\star \simeq 4 \etap c_s / \omegalin \simeq (4\etap / 6.223) c_s f_\textrm{max}$ \cite{Witkowski:2021raz, Fumagalli:2021mpc}, with $c_s$ the propagation speed of scalar fluctuations when the post-inflationary GWs are induced. The inflationary-era GW spectrum is well-approximated by the template \eqref{eq:GW-inf-template} for $f < 4 \etap / \omegalin$, which we implemented explicitly by including a Heaviside theta function. The sinusoidal terms give rise to oscillations in the GW spectrum for $f > f_\textrm{max}$. These oscillations occur with the same frequency $\omegalin$ that also appears in the scalar power spectrum \eqref{eq:P-sharp-intro} and they can be checked to have unit amplitude. Note that in contrast to the post-inflationary contribution there is no model-dependent envelope, so that the functional form of \eqref{eq:GW-inf-template} is robust for a wide range of realisations of a sharp feature. The model parameters for this template are thus $(h^2 \ObarGW, \, \etap, \, \omegalin)$. 
Both the templates \eqref{eq:GW-sharp-template} and \eqref{eq:GW-inf-template} describe contributions to $h^2 \OGW$ due to the same underlying physical process, a sharp feature that occurred during inflation. As a result, the parameters appearing in both templates are related as they can be expressed in terms of the physical parameters that describe the sharp feature.\footnote{Even though we use the same label $h^2 \ObarGW$ as a parameter for the overall amplitude in \eqref{eq:GW-sharp-template} and \eqref{eq:GW-inf-template}, this does not imply that this parameter has to take the same value in \eqref{eq:GW-sharp-template} and \eqref{eq:GW-inf-template} for the same sharp feature. The relation between $h^2 \ObarGW^{\textrm{inf.-era}}$ and $h^2 \ObarGW^{\textrm{post-inf.}}$ is more involved and also depends on other model parameters, see \cite{Fumagalli:2021mpc} for more details.} 
However, these relations will not be crucial for this work. Here we will assume that only one of the two signals \eqref{eq:GW-sharp-template} or \eqref{eq:GW-inf-template} is detectable by LISA. This occurs when either \eqref{eq:GW-sharp-template} or \eqref{eq:GW-inf-template} dominates over the other contribution. In addition, even if these two contributions have a similar amplitude, LISA may still only be sensitive to one of them, as the templates \eqref{eq:GW-sharp-template} and \eqref{eq:GW-inf-template} peak at a different value of $f$. Thus, for the template \eqref{eq:GW-inf-template} we will treat $(h^2 \ObarGW, \, \etap, \, \omegalin)$ as effectively free parameters.
This does not imply that one should exclude the possibility that both \eqref{eq:GW-sharp-template} and \eqref{eq:GW-inf-template} are simultaneously detectable by LISA, which can occur in certain regions of parameter space. This would offer a unique opportunity for extracting information about the underlying sharp feature. Here, we leave this exciting possibility for future work.\footnote{To consider this case one would also need to include the mixed terms in $h^2 \OGW(f)$ from contributions of type $\sim \langle h^{\textrm{inf.-era}} h^{\textrm{post-inf.}} \rangle$, which we have ignored so far. However, we expect these mixed terms to have little effect on the signal where \eqref{eq:GW-sharp-template} and \eqref{eq:GW-inf-template} respectively peak, as the respective other contribution is small there, which should also suppress the mixed piece.}  

It behoves to point out that the template \eqref{eq:GW-inf-template} is not only valid for sharp features, but can arise in more general contexts. In fact, an inflationary-era induced GW spectrum of the form \eqref{eq:GW-inf-template} is associated with dynamically generated excited states for scalar fluctuations during inflation \cite{Fumagalli:2021mpc}, and sharp features are only one possible mechanism for generating this. The expression \eqref{eq:GW-inf-template} is valid if the excited state exhibits copious particle production and if only a narrow band of scales is affected by it, as is always the case if the excited state is `prepared' by a sharp feature that at the same time enhances $\Pzeta$. If a broader range of scales is in an excited state, the shapes of the peak and of the UV tail of the GW spectrum change, but the presence of oscillations with frequency $\omegalin$ on the UV tail is robust, see again \cite{Fumagalli:2021mpc}.    

\begin{figure}[t]
\centering
\begin{overpic}[width=0.7\textwidth]{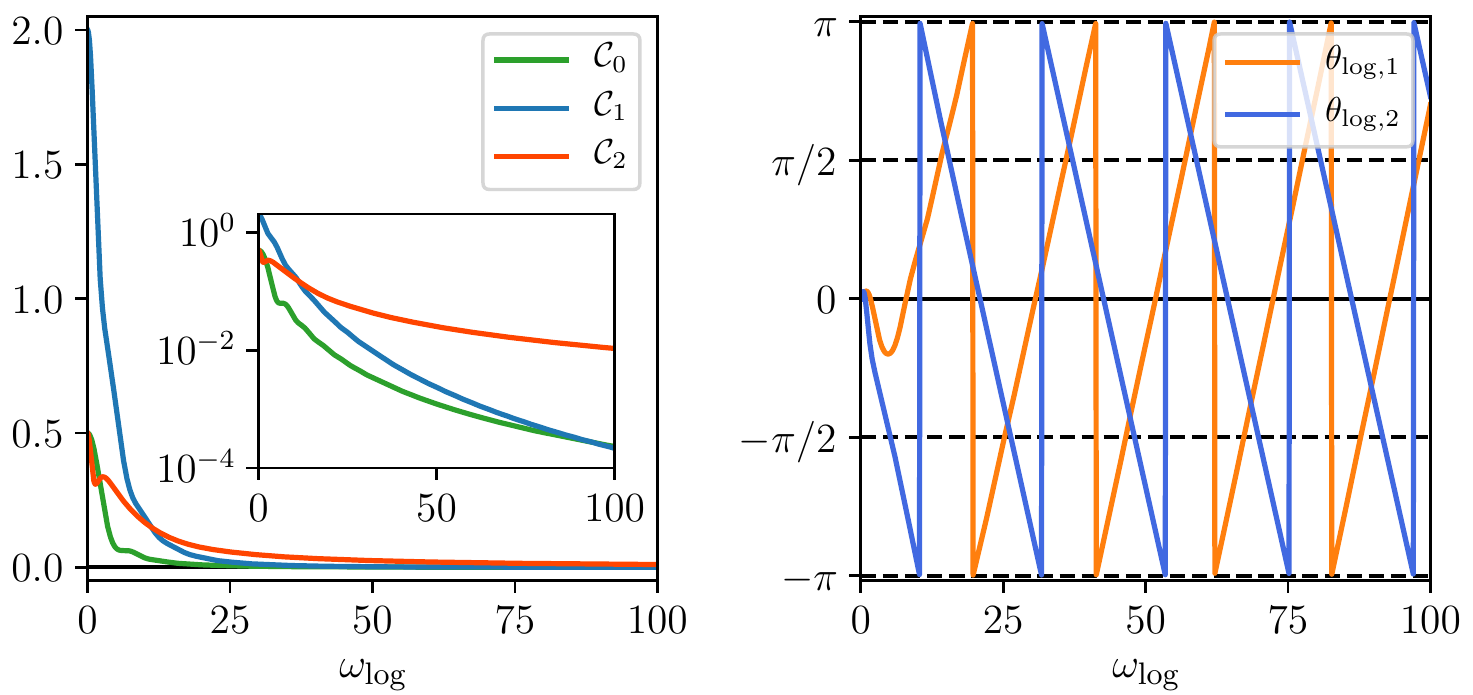}
\end{overpic}
\caption{Plots of the functions $\C_{0,1,2}(\omegalog)$ and $\theta_{\textrm{log},1,2}(\omegalog)$ that appear in the template \eqref{eq:resonant-template-A} for the post-inflationary contribution to the SGWB due to a resonant feature. The expressions for computing these functions are collected in appendix \ref{app:resonant-template-coeffs}. The plots shown here are obtained for GWs sourced during a period of radiation domination.}
\label{fig:C012theta12}
\end{figure}

\subsection{Resonant feature: post-inflationary contribution}
\label{sec:templates-resonant-post}
The post-inflationary contribution to the GW spectrum from a resonant feature during inflation follows the template \eqref{eq:resonant-template-intro}, i.e.~the oscillatory part is given by a superposition of two sinusoidal pieces with frequencies $\omegalog$ and $2 \omegalog$, respectively, that modulate a common envelope $\ObarGW$. There are cross-correlations between the parameters $\mathcal{A}_{\textrm{log},1,2}$, $\theta_{\textrm{log},1,2}$, the frequency $\omegalog$ and the shape of the envelope \cite{Fumagalli:2021cel,Witkowski:2021raz}. For example, for small $\omegalog$ the amplitude $\mathcal{A}_{\textrm{log},1}$ is significantly larger than $\mathcal{A}_{\textrm{log},2}$ for general envelopes, while for sufficiently large $\omegalog$ this hierarchy is inverted, while both $\mathcal{A}_{\textrm{log},1,2}$ decrease with increasing $\omegalog$. For envelopes that are effectively constant over several periods of oscillation, the parameters $\mathcal{A}_{\textrm{log},1,2}$ and $\theta_{\textrm{log},1,2}$ can be computed exactly as a function of $\omegalog$ \cite{Fumagalli:2021cel}. The result also depends on the equation of state of the Universe when the GWs are induced \cite{Witkowski:2021raz}.

In this work we are interested in the question of to what extent a representative example of a resonant feature can be detected with LISA. Hence we find it important to take into account realistic relations between the parameters $\mathcal{A}_{\textrm{log},1,2}$, $\theta_{\textrm{log},1,2}$ and $\omegalog$. Thus, here we will consider the case where the envelope $\ObarGW$ is constant, at least over the range of frequencies that LISA is sensitive to. Using the results from \cite{Fumagalli:2021cel}, we can then re-express the parameters $\mathcal{A}_{\textrm{log},1,2}$ and $\theta_{\textrm{log},1,2}$ directly in terms of the parameters $A_\textrm{log}$ and $\omegalog$ appearing in \eqref{eq:P-resonant-intro}. The template \eqref{eq:resonant-template-intro} becomes:
\begin{align}
    \label{eq:resonant-template-A} 
    h^2 \Omega_{\textrm{GW}}(f) = h^2 \overline{\Omega}_{\textrm{GW}} \bigg[1 &+ \frac{A_{\textrm{log}} \C_1(\omegalog)}{1 + A_{\textrm{log}}^2 \C_0(\omegalog)} \cos \big(\omegalog \ln (f/ \textrm{Hz}) + \philog+ \theta_{\textrm{log},1}(\omegalog) \big) \\
    \nonumber & + \frac{A_{\textrm{log}}^2 \C_2(\omegalog)}{1 + A_{\textrm{log}}^2 \C_0(\omegalog)} \cos \big(2 \omegalog \ln (f/ \textrm{Hz}) + 2 \philog + \theta_{\textrm{log},2}(\omegalog) \big) \bigg] \, ,
\end{align}
where $\C_{0,1,2}(\omegalog)$ and $\theta_{\textrm{log},1,2}(\omegalog)$ can be computed numerically using the expressions in appendix \ref{app:resonant-template-coeffs} and we have set $\fref = 1 \textrm{ Hz}$ for simplicity.\footnote{The piece $ A_\textrm{log}^2 \C_0$ is a constant contribution to $h^2 \OGW$ due to the oscillations in $\Pzeta$, while $A_\textrm{log} \C_1$ and $A_\textrm{log}^2 \C_2$ set the amplitudes of the two oscillatory terms. Here we factored out the constant piece $(1+A_\textrm{log}^2 \C_0)$ and absorbed it into $h^2 \ObarGW$, resulting in this term appearing in the denominator of the amplitudes multiplying the oscillatory pieces.} Here we will restrict our attention to GWs induced during a period of radiation domination and in fig.~\ref{fig:C012theta12} plot the functions $\C_{0,1,2}(\omegalog)$ and $\theta_{\textrm{log},1,2}(\omegalog)$ for this case. For a different equation of state during GW production, the functions $\C_{0,1,2}(\omegalog)$ and $\theta_{\textrm{log},1,2}(\omegalog)$ differ quantitatively, but the qualitative behaviour with $\omegalog$ is similar \cite{Witkowski:2021raz}. The phase $\philog$ in \eqref{eq:P-resonant-intro} also contributes to the phase in the GW spectrum, and here we separated this dependence from $\theta_{\textrm{log},1,2}(\omegalog)$. The model parameters for this template are hence ($h^2\ObarGW, \, A_\textrm{log}, \omegalog, \philog$). In the right panel of fig.~\ref{fig:inf_sharp_res_post_examples} we show two examples of the template \eqref{eq:resonant-template-A} (see the caption for the model parameters). 

\section{Principal Component Analysis}
\label{sec:PCA}
As a first step to test the capability of LISA of probing the presence of features in the SGWB, we apply the PCA analysis of~\cite{Pieroni:2020rob} to some mock data generated with procedure analogous to the one employed in~\cite{Caprini:2019pxz, Pieroni:2020rob, Flauger:2020qyi}. Namely, we start by dividing the full observation time of LISA (\emph{i.e.}~4 years with 75$\%$ efficiency) into $N_c$ data segments with a duration of around $11.5$ days, giving a frequency resolution of around $10^{-6}$Hz. For each of these data segments we generate a Gaussian realization of both signal and noise\footnote{We employ the same noise model as~\cite{Flauger:2020qyi}, we factor out the detector's response function and we convert to $\Omega$ units to directly work with the signals defined as in~\cref{sec:templates}.} and, after self-correlating the data with themselves, we average over segments. After this procedure, we coarse-grain the dataset following the procedure described in Sec.~3.1 of~\cite{Flauger:2020qyi} to reduce the numerical complexity of the problem. In this section we will restrict our analysis to a single data stream, corresponding to the TDI X channel, and we will denote the data points as $D_k$, where the index $k$ runs over the reduced set of frequencies $f_k$, and their weights\footnote{Which simply correspond to the number of points we averaged over in the coarse-graining procedure.} with $n_k$. In this work we neglect the effects of astrophysical foregrounds of either galactic or extra-galactic origin~\cite{Bender:1997hs, Farmer:2003pa, Barack:2004wc, Regimbau:2009rk, Regimbau:2011rp, LIGOScientific:2019vic, Lamberts:2019nyk, KAGRA:2021kbb} which require more elaborate techniques~\cite{Karnesis:2021tsh}. For completeness, the SNR is computed as:\footnote{Or with the integral replaced by the corresponding sum over the discrete frequencies $f_k$.}
\begin{align}
\label{eq:SNRdef}
    \textrm{SNR} = \sqrt{T \int_0^\infty \textrm{d}f \left( \frac{S^{\rm th}}{N^{\rm th}} \right)^2} \;,
\end{align}
where $T$ is the total effective observation time, $S^{\rm th}(f_k, \vec{\theta}_s)$ denotes the theoretical model for the signal, defined in terms of a set of parameters $\vec{\theta}_s$ and $N^{\rm th}(f_k, \vec{\theta}_n)$ is the noise model, defined in terms of $\vec{\theta}_n$. In practice, $S^{\rm th}$ and $N^{\rm th}$ are the fiducial models for the signal and the LISA sensitivity in units of $h^2 \OGW$.  

In the following, we assume the data, as defined in the previous paragraph, to be described by a Gaussian likelihood of form:
\begin{equation}
    \label{eq:Gaussian_lik}
		- \ln \mathcal{L} (\vec{\theta}_s, \vec{\theta}_n) \simeq \frac{N_c}{2} \sum_k n_k \left[   \frac{ D_k - S^{\rm th}(f_k, \vec{\theta}_s) - N^{\rm th}(f_k, \vec{\theta}_n) }{ D_k} \right]^2 \; ,
	\end{equation}
where we have neglected irrelevant normalization factors. In the following we denote with $\vec{\theta}$ the whole set of parameters running over both $\vec{\theta}_s$ and $\vec{\theta}_n$. This likelihood is known to be low-biased~\cite{Bond:1998qg, Sievers:2002tq, WMAP:2003pyh, Hamimeche:2008ai}. A more accurate description of these data can be obtained by introducing an additional Lognormal contribution to the likelihood~\cite{WMAP:2003pyh} which allows for a better modeling of the skewness contribution, reducing the bias in the estimate of the $\vec{\theta}$. However, since this low-bias is irrelevant for the purpose of the present analysis (which only aims at recognizing the frequency shape of the SGWB), and in order to provide a likelihood which is quadratic in the model parameters~\footnote{The reason for this choice will be clarified in the following.}, we stick to the Gaussian approximation given in~\cref{eq:Gaussian_lik}. 

\begin{figure}
    \centering
    \includegraphics[width=.7\textwidth]{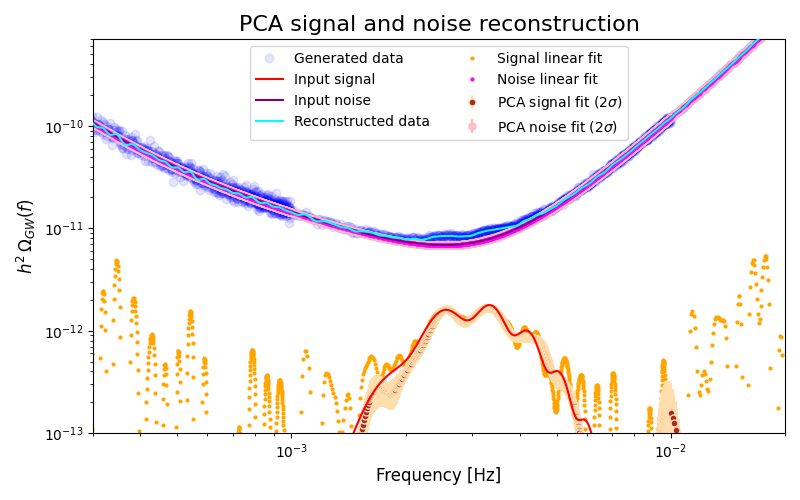}
    \caption{PCA reconstruction of a sharp feature of type \eqref{eq:GW-sharp-template} with lognormal envelope \eqref{eq:lognorm-env} with model parameters as given in the main text of sec.~\ref{sec:PCA}. The signal and noise `linear fits' correspond to the solutions \eqref{eq:Maximum_likelihood} for the signal parameters $\theta_s^j$ in \eqref{eq:signal_model} and the corresponding noise parameters. The PCA reconstruction of the signal is shown as the `PCA signal fit' together with its $2\sigma$ band.  This shows that the sharp feature type oscillations near the peak can be successfully reconstructed.}
    \label{fig:PCA_plots_sharp_lognorm}
\end{figure}

\begin{figure}
    \centering
    \includegraphics[width=.7\textwidth]{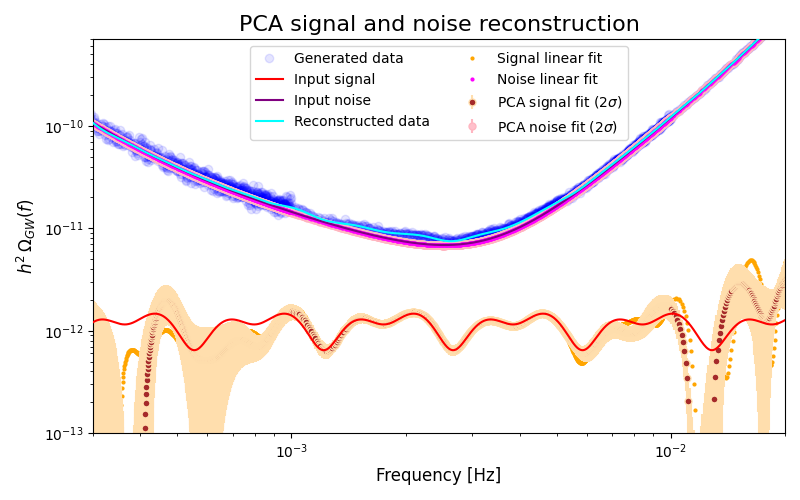}
    \caption{As fig.~\ref{fig:PCA_plots_sharp_lognorm} but for a resonant feature of type \eqref{eq:resonant-template-A} with model parameters as given in the main text of sec.~\ref{sec:PCA}. Here the `signal linear fit' curve is largely obscured by the $2\sigma$ band of the `PCA signal fit' curve. The signal is successfully reconstructed for two periods allowing for an identification of the resonant feature type oscillation.}
    \label{fig:PCA_plots_resonant_flat}
\end{figure}

Assuming $S^{\rm th}(f_k, \vec{\theta}_s)$ and $N^{\rm th}(f_k, \vec{\theta}_n)$ are linear in $\vec{\theta}_s$ and $\vec{\theta}_n$ respectively, the theoretical model for the data can be expressed as $D^{\rm th}(f_k, \vec{\theta}) = S^{\rm th}(f_k, \vec{\theta}_s) + N^{\rm th}(f_k, \vec{\theta}_n) = A_{i} \theta^i$. Thus, the problem of determining the maximum likelihood estimate for the $\vec{\theta}$ can be cast as a linear problem which is solved by:
\begin{equation}
    \label{eq:Maximum_likelihood}
    - \frac{ \partial \ln \mathcal{L} }{\partial \theta^i} = 0 \; , \qquad  \longrightarrow  \qquad   \bar{ \theta }^i = F^{-1}_{ij} N_c \sum_k \frac{n_k}{D_k} A_j   \; ,
\end{equation}
where $F^{-1}_{ij}$ denotes the inverse of the Fisher Information Matrix (FIM), \emph{i.e.}~the covariance matrix, defined as:
\begin{equation}
    \label{eq:fisher_matrix}
   F_{ij} \equiv \left. \frac{ \partial^2 \ln \mathcal{L} }{\partial \theta^i \partial \theta^j} \right|_{\theta^i = \bar{\theta}^i }= \left. N_c \sum_k \frac{n_k}{D_k^2} \frac{ \partial D^{\rm th}(f_k, \vec{\theta})}{\partial \theta^i } \frac{ \partial D^{\rm th}(f_k, \vec{\theta})}{\partial \theta^j }   \right|_{\theta^i = \bar{\theta}^i } \; ,
 \end{equation}
 and $\bar{\theta}^i$ are the best-fit parameters. Under the assumption of a model which is linear in the parameters, it is thus easy to show that the FIM reduces to:
\begin{equation}
    \label{eq:fisher_matrix_PCA}
   F_{ij} = N_c \sum_k \frac{n_k}{D_k^2} A_i A_j \; , 
\end{equation}
which does not depend on the model parameters. Up to this point we have not specified any particular model for the signal and for the noise. Concerning noise we use the model described in~\cite{Pieroni:2020rob}, which is equivalent to the noise model of~\cite{Flauger:2020qyi, Babak:2021mhe} with one caveat: we redefine the noise parameters in order to have a model which is linear in $\vec{\theta}_n$ and the new noise parameters are equal to one at face value. Concerning signal, in the rest of this section we assume the signal can be expanded as a sum over $m$ normalized Gaussians $\delta_\sigma$ (with zero mean and some $\sigma $) each centered around a certain pivot frequency $f_j$ with $j \in 1, ..., m$ as:
\begin{equation}
    \label{eq:signal_model}
    S^{\rm th}(f_k, \vec{\theta}_s) = \sum_{j=1}^m \theta_s^j \;  \delta_\sigma (f_k -f_j) \; , 
\end{equation}
where $\theta_s^j$ are the parameters\footnote{As in~\cite{Pieroni:2020rob}, the $\sigma$ is kept fixed in the analysis. In a fully consistent Bayesian framework this parameter would have to be determined together with the $\theta_s^j$. However, since for the scopes of this work we are only interested in checking whether the features discussed in~\cref{sec:templates} could be identified, we simply fixed this quantity to a reasonable value which allows to resolve the features. A similar argument holds for the choice of the pivot frequencies $f_j$.} to be determined through~\cref{eq:Maximum_likelihood}. As described in~\cite{Pieroni:2020rob}, the idea of PCA is to decompose the signal in terms of the well-determined components only. For this purpose, we start by computing the eigensystem $\{ \lambda_k$, $e^{(k)}_i \}$, where $\lambda_k$ are the eigenvalues, and $e^{(k)}_i$ are the corresponding eigenvectors, of the FIM in~\cref{eq:fisher_matrix_PCA}. We then project the signal and the noise onto the basis defined by the $e^{(k)}_i$ and we drop all the components corresponding to parameters which are compatible with zero\footnote{Notice that in this basis the errors are just given by $\lambda_k^{-1/2}$. } at $2\sigma$. Finally, we express both signal and noise in terms of this cut basis.

\begin{figure}
    \centering
    \includegraphics[width=.70\textwidth]{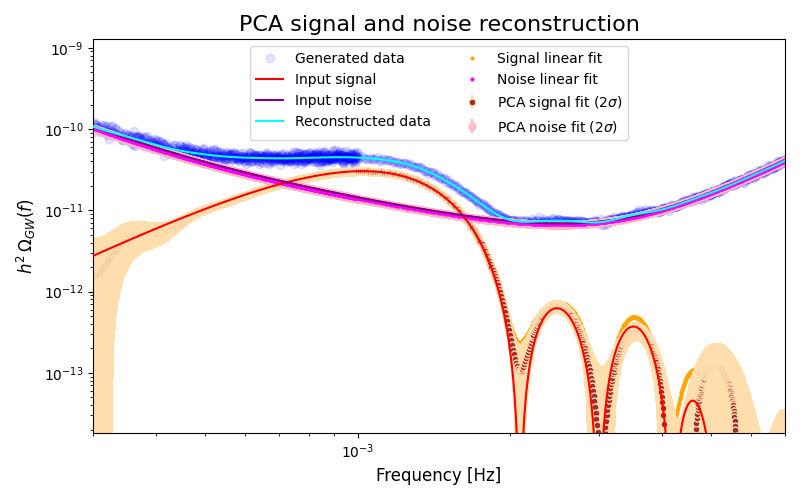}
    \caption{As fig.~\ref{fig:PCA_plots_sharp_lognorm} but for the inflationary-era contribution due to a sharp feature in \eqref{eq:GW-inf-template} with model parameters as given in the main text of sec.~\ref{sec:PCA}. Here the `signal linear fit' curve is largely obscured by the $2\sigma$ band of the `PCA signal fit' curve. In addition to the principal peak, the PCA analysis succeeds in reconstructing the first two oscillations on the UV tail.}
    \label{fig:PCA_plots_sharp_inf}
\end{figure}

In figs.~\ref{fig:PCA_plots_sharp_lognorm}, \ref{fig:PCA_plots_resonant_flat} and \ref{fig:PCA_plots_sharp_inf} we show the result of the PCA procedure on three mock signals: a sharp feature with lognormal envelope as in~\cref{eq:GW-sharp-template} with \eqref{eq:lognorm-env}, a resonant feature with flat envelope as in~\cref{eq:resonant-template-A} and finally the inflationary-era contribution due to a sharp feature~as in \cref{eq:GW-inf-template}. In particular the parameters for the three models are respectively:
\begin{itemize}
    \item Lognormal sharp (SNR $\simeq 80$): $\log_{\rm 10} ( h^2 \ObarGW) = -11.8$, $f_{\star} = 3$ mHz, $\Delta = 0.3$, with the parameters describing the oscillation given by $\mathcal{A}_{\rm lin} = 0.2$, $\omegagwlin = 7 $ mHz${}^{-1}$ and $\theta_{\rm lin} = \pi /2 $. For this analysis we used $150$ evenly logarithmically spaced Gaussians with width $\sigma_j =0.05 \, f_j$. The model parameters satisfy $\omegalin f_\star \Delta \approx 2 \pi$ which is consistent with choosing the relatively large oscillation amplitude $\mathcal{A}_{\rm lin} = 0.2$, see the discussion in sec.~\ref{sec:templates-sharp-post}.
    \item Flat resonant (SNR $\simeq 85$): $\log_{\rm 10} ( h^2 \ObarGW) = -11.95$ with resonant feature characterized by $A_{\rm log} = 1$, $\omega_{\rm log} = 8 $ and $\phi_{\rm log} = 0$. In this analysis we used $50$ evenly logarithmically spaced Gaussians with width $\sigma_j =0.1 \, f_j$. For this choice of parameters the two oscillatory pieces in \eqref{eq:resonant-template-A} have a comparable amplitude, leading to the intricate oscillatory pattern. 
    \item Inflationary sharp (SNR $\simeq 562$): $\log_{\rm 10} ( h^2 \ObarGW) = -10.5$, $\gamma = 10$ and $\omega_{\rm lin} = 6 \textrm{ mHz}^{-1}$. In this analysis we used $50$ evenly logarithmically spaced Gaussians with width $\sigma_j =0.5 \, f_j$. As the oscillations only occur on the UV tail of this template, oscillations visible to LISA necessarily come together with a high-amplitude main peak and a large SNR.  
\end{itemize}
As one can see from the three plots in figs.~\ref{fig:PCA_plots_sharp_lognorm}, \ref{fig:PCA_plots_resonant_flat} and \ref{fig:PCA_plots_sharp_inf}, after dropping the low information components (which are still present in the orange dots in these plots), the signal can be quite well reconstructed (brown dots with navajo-shaded error bands) and the features can be reasonably well identified. Assuming that the right template to describe the signal component is identified, it would be possible to forecast the constrains on the model parameters. This will be the topic of the next section, where this procedure will be performed using the FIM to define a Gaussian approximation of the Likelihood around the best fit.

\section{Fisher forecast}
\label{sec:Fisher}

 In this next section we derive the FIM forecasts on some of the parameters of the three templates presented in \eqref{eq:GW-sharp-template}, \eqref{eq:GW-inf-template} and \eqref{eq:resonant-template-A}. After a short review of the basics of the FIM formalism, we specialize our analysis to these three cases. 
 
\subsection{Setting up a Fisher forecasts for LISA}
The idea of the FIM formalism is to assume that the likelihood can be expanded up to quadratic order in the model parameters $\theta^{i}$ (with $i \in 1, .. , k$, where $k$ is the number of parameters) around the best fit $\bar{\theta}^{i}$ as:\footnote{In order to produce the forecasts it is thus necessary to assume that the best-fit parameters $\bar{\theta}^{i}$ obtained by directly evaluating the likelihood match exactly the parameters of the injected signal.}
\begin{equation}
    -2 \ln \mathcal{L} (\vec{\theta}) =  \ln\left[ (2 \pi)^k / \textrm{det}(F_{i j}) \right] + (\theta^{i} -\bar{\theta}^{i}) F_{i j} (\theta^{j} -\bar{\theta}^{j})  \; ,
\end{equation}
where $F_{ij}$ is the FIM defined as in~\cref{eq:fisher_matrix} upon replacing $D_k$ with $D^{\rm th}(f_k, \bar{\theta}^{i})$, \emph{i.e}.~by assuming that the data match exactly the fiducial model. Within this framework, the covariance matrix $C_{i j}$ can be directly expressed as $C_{i j} = F_{i j}^{-1}$ and the confidence intervals for the model parameters can directly be computed, without directly evaluating/sampling the parameter space of the likelihood. Despite being very effective while being extremely simple (the $1$ and $2\sigma$ contours can be computed analytically from $C_{i j}$), this approach presents some limitations. For example, within this framework it is not possible to properly answer the question of identifying the best-suited fiducial model to describe the dataset.

In this section $S^{\rm th}(f_k, \vec{\theta}_s)$ is identified with one of the templates for $h^2 \OGW$ in \eqref{eq:GW-sharp-template}, \eqref{eq:GW-inf-template} and \eqref{eq:resonant-template-A}, with $\vec{\theta}_s$ the parameters appearing in these templates. For the sake of simplicity, in the analyses presented in the rest of this section we also work under the assumption that the noise parameters are known \emph{i.e.}~the noise spectrum is included in $D^{\rm th}(f_k, \bar{\theta}^{i})$ but we don't include the noise parameters in the FIM. This corresponds to assuming the noise parameters to be known, which leads to an optimistic determination of the constraints. A more comprehensive examination of these models, where the noise parameters will also be included in the analyses is left to future works on topic. Similarly, we stress once again that we ignore the impact of foregrounds on the reconstruction of the signal parameters. These additional complications are also postponed to future studies. 

\subsection{Sharp feature: post-inflationary contribution}
\label{sec:fisher-sharp-post}
\subsubsection{Flat envelope}
\label{sec:fisher-sharp-post-flat}
\begin{figure}[t]
\centering
\begin{overpic}[width=0.65\textwidth]{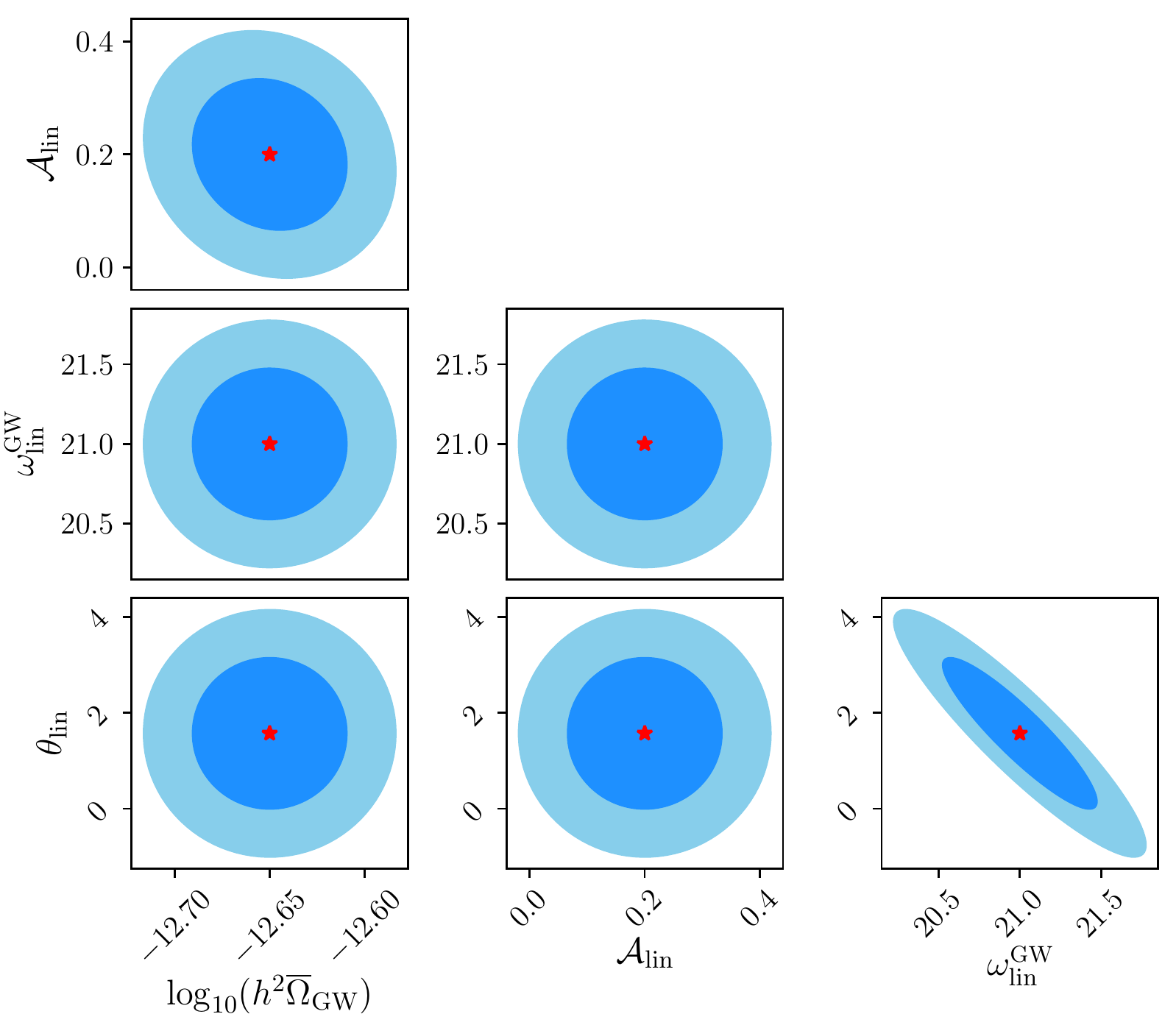}
\end{overpic}
\caption{Confidence ellipses with $1\sigma$ and $2\sigma$ regions obtained from a Fisher forecast for the post-inflationary contribution to the SGWB due to a sharp feature with template \eqref{eq:GW-sharp-template} with flat envelope \eqref{eq:flat-env}. The model parameters are given in \eqref{eq:sharp-post-benchmark-I}. The dimensionful quantity $\omegagwlin$ is in units of $\textrm{mHz}^{-1}$. The SNR for this model is 16.8. Note that $\mathcal{A}_\textrm{lin}=0$ is just within the $2\sigma$ contour implying that the oscillation is close to the limit of detectability in this case.}
\label{fig:corner-flat-sharp}
\end{figure}

We begin with the post-inflationary contribution to the SGWB due to a sharp feature with a flat envelope, i.e.~for $h^2 \OGW$ as given in \eqref{eq:GW-sharp-template} with \eqref{eq:flat-env}. This template signal, while maybe not very realistic due to the flat envelope, will exhibit the main issues regarding the detectability of feature contributions to the SGWB with LISA. The flat envelope example is also interesting as it represents the most optimistic scenario for detecting a sharp feature type oscillation. If an oscillating signal with a flat envelope is found to be undetectable by LISA, we expect any peaked signal with the same value of $h^2 \ObarGW$ and modulated by the same type of oscillation to also be undetectable. In fig.~\ref{fig:corner-flat-sharp} we plot the confidence ellipses for the $1\sigma$ and $2\sigma$ contours for all pairwise combinations of the template parameters $(h^2 \ObarGW, \, \mathcal{A}_\textrm{lin}, \, \omegagwlin, \, \theta_\textrm{lin})$ for a model with parameters values:
\begin{align}
\label{eq:sharp-post-benchmark-I}
\log_{10} (h^2 \ObarGW) =-12.65, \quad \mathcal{A}_\textrm{lin} = 20 \%, \quad \omegagwlin = 21 \textrm{ mHz}^{-1}, \quad \theta_\textrm{lin}=\frac{\pi}{2} \, .
\end{align}
The value $\mathcal{A}_\textrm{lin} = 20 \%$ is chosen as this is close to the maximum that is expected for post-inflationary GWs due to a sharp feature. The choice $\omegagwlin = 21 \textrm{ mHz}^{-1}$ was made for later convenience.\footnote{We will later also consider GW spectra with a peaked envelope that fall into the region of maximal sensitivity of LISA around $f \simeq 3 \textrm{ mHz}$. A frequency $\omegagwlin = \mathcal{O}(10) \textrm{ mHz}^{-1}$ will then be consistent with an $\mathcal{O}(1)$ number of oscillations modulating this peak, consistent with a detectable amplitude $\mathcal{A}_\textrm{lin} = \mathcal{O}(10) \%$, see the discussion at the end of sec.~\ref{sec:templates-sharp-post}.} A plot of $h^2 \OGW(f)$ for these parameter values is shown in fig.~\ref{fig:sharp_post_examples} (blue curve). 

We extract the following information from fig.~\ref{fig:corner-flat-sharp}. All ellipses except the one in the $\theta_\textrm{lin}$-$\omegagwlin$-panel do not have a significant tilt of their major axis, implying little correlation between the parameters except between the frequency of the oscillation and the phase. We understand this to be due to the fact that the most sensitive frequency range of LISA only has access to a few periods of oscillation. As a result, an error in the value of $\omegagwlin$ can be compensated by a shift in $\theta_\textrm{lin}$. The small tilt of the ellipse axes in the $\mathcal{A}_\textrm{lin}$-$\log_{10} (h^2 \ObarGW)$-panel implies that here an underestimate of $\log_{10} (h^2 \ObarGW)$ weakly correlates with an overestimate of $\mathcal{A}_\textrm{lin}$ and vice-versa. 

The subject of main interest is, to what extent the oscillation in the GW signal is detectable with LISA and hence we now focus on the parameters $\mathcal{A}_\textrm{lin}$ and $\omegagwlin$ characterising the oscillation.\footnote{We ignore $\theta_\textrm{lin}$ whose precise value is not important for the existence of an oscillation.} From the covariance matrix, the errors\footnote{The error for a parameter $x$ is given here by the square root of the corresponding diagonal element of the covariance matrix, i.e.~$\sigma(x)=\sqrt{C_{xx}}$.} for these two parameters can be deduced as $\sigma(\mathcal{A}_\textrm{lin}) \simeq 0.088$ and $\sigma(\omegagwlin) \simeq 0.31 \textrm{ mHz}^{-1}$ for the example in fig.~\ref{fig:corner-flat-sharp}. For $\omegagwlin$ this translates into a small relative error of $\sim 1.5 \%$, while for $\mathcal{A}_\textrm{lin}$ the relative error is $\sim 44 \%$. In a related manner, in fig.~\ref{fig:corner-flat-sharp}, the value $\mathcal{A}_\textrm{lin}=0$ can be seen to just fall within the $2\sigma$ contour. Thus, for the model parameters chosen here, the Fisher forecast suggests that the oscillating signal is close to the edge of detectability with LISA: e.g.~in a blind reconstruction of this signal in LISA we hence expect a significant chance of misidentifying it as a smooth signal due to an inability of sufficiently resolving the amplitude of oscillation $\mathcal{A}_\textrm{lin}$. Also note that the SNR is fairly low for this example model, with $\textrm{SNR}=16.8$.

By considering other parameter choices than those in \eqref{eq:sharp-post-benchmark-I} one can check that it is in general the error on $\mathcal{A}_\textrm{lin}$ that appears to limit the detectability of an oscillating signal of type \eqref{eq:GW-sharp-template} in LISA. If we consider $\omegagwlin > \mathcal{O}(10) \textrm{ mHz}^{-1}$ (and all other parameters unchanged), the absolute errors on $\mathcal{A}_\textrm{lin}$ and $\omegagwlin$ do not change by much. For $\omegagwlin < \mathcal{O}(1) \textrm{ mHz}^{-1}$ the error on $\omegagwlin$ increases, but so does the error on $\mathcal{A}_\textrm{lin}$.\footnote{This is due to the fact that for $\omegagwlin < \mathcal{O}(1) \textrm{ mHz}^{-1}$ there is in general less than one period of oscillation within the frequency band of maximal LISA sensitivity.} The value of $\mathcal{A}_\textrm{lin}$ itself does not affect the absolute error very much and the value $\mathcal{A}_\textrm{lin} = 20 \%$ is already close to the maximal value for GWs induced during a post-inflationary universe dominated by radiation. While for a post-inflationary era with a stiffer equation of state values up to $\mathcal{A}_\textrm{lin} = 40 \%$ can occur, see \cite{Witkowski:2021raz}, these maximal values are only realised for particular parameter choices, while values $\mathcal{A}_\textrm{lin} \sim 10 \%$ are much more common. Unsurprisingly, what has the most impact on the error of $\mathcal{A}_\textrm{lin}$ (and all other model parameters) is the value of the absolute amplitude $h^2\ObarGW$. Increasing $h^2\ObarGW$ the errors on the parameters decrease, while they increase as $h^2\ObarGW$ is lowered. In the latter case the SNR will also very quickly approach unity. To give a few examples, in the table below we list the relative error in $\mathcal{A}_\textrm{lin}$ and the SNR for various values of $h^2\ObarGW$ for the flat envelope sharp feature with all other parameters as in \eqref{eq:sharp-post-benchmark-I}:
\begin{center}
\small
\begin{tabular}{| l || c | c | c | c | c | c | c | c |} \hline 
  \cellcolor[gray]{0.95} $\log_{10}(h^2\ObarGW)$ & -13 & -12.65 & -12.5 & -12.25 & -12 & -11.5 & -11 & -10 \\ \hline
  \cellcolor[gray]{0.95} $\sigma(\mathcal{A}_\textrm{lin}) / \mathcal{A}_\textrm{lin}$ & 97\% & 44\% & 31\% & 18\% & 10\% & 3.8\% & 1.7\% & 0.7\% \\ \hline
  \cellcolor[gray]{0.95} SNR & 7.5 & 16.8 & 23.7 & 42.2 & 75.0 & 237 & 750 & 7500 \\ \hline
\end{tabular}
\end{center}
Thus, for $h^2 \ObarGW \lesssim 10^{-13}$ we expect that the signal may not be detectable as the SNR becomes too small. For $h^2 \ObarGW \lesssim 10^{-12.5}$ the signal may be still detectable, but the oscillations may not be observable as suggested by the error on $\mathcal{A}_\textrm{lin}$. Note that the SNR grows exactly linearly with the quantity $h^2 \ObarGW$. This follows from the definition of the SNR in \eqref{eq:SNRdef} and the fact that $h^2 \ObarGW$ appears as an overall multiplicative factor in the template. This is also the case for all other templates considered in this work, and hence this relation holds in general. 

\begin{figure}[t]
\centering
\begin{overpic}[width=0.98\textwidth]{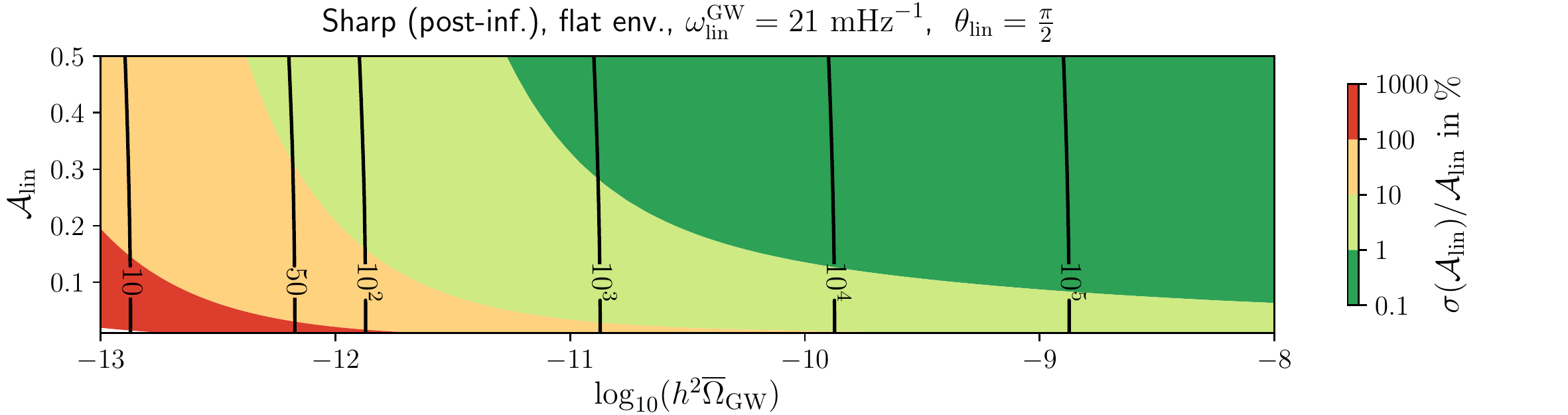}
\end{overpic}
\caption{Relative error $\sigma(\mathcal{A}_\textrm{lin}) / \mathcal{A}_\textrm{lin}$ in $\%$ from a Fisher forecast for a sharp feature with template \eqref{eq:GW-sharp-template} and a flat envelope, plotted as a function of $h^2 \OGW(f)$ and $\mathcal{A}_\textrm{lin}$. The remaining model parameters are as in \eqref{eq:sharp-post-benchmark-I}. The black contours denote lines of constant SNR.}
\label{fig:sharp_post_flat_Alin_logObar_21_pio2}
\end{figure}

\begin{figure}[t]
\centering
\begin{overpic}[width=0.98\textwidth]{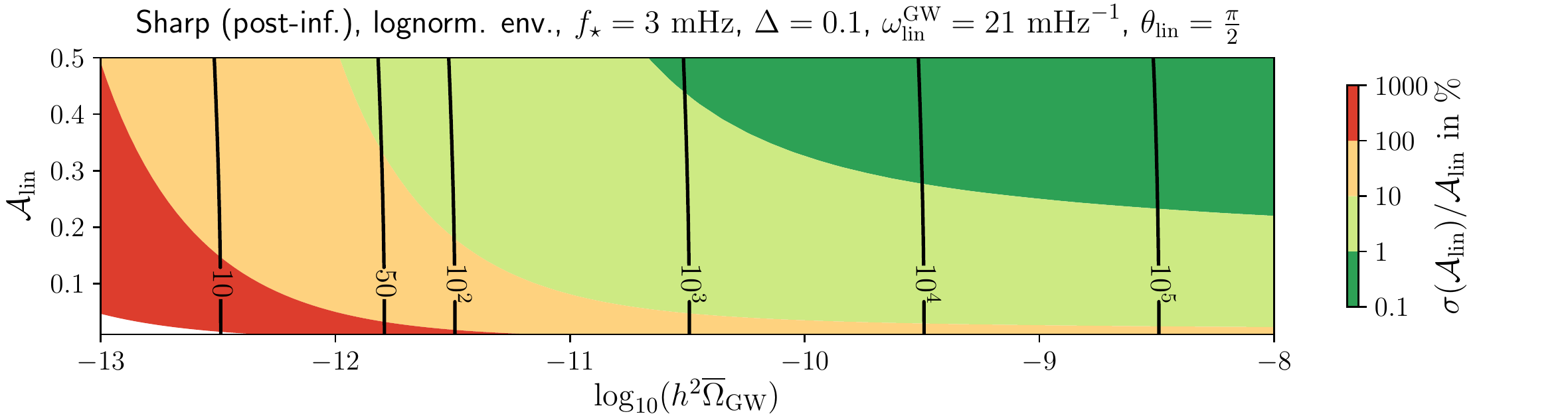}
\end{overpic}
\caption{Relative error $\sigma(\mathcal{A}_\textrm{lin}) / \mathcal{A}_\textrm{lin}$ in $\%$ from a Fisher forecast for a sharp feature with template \eqref{eq:GW-sharp-template} and a lognormal envelope, plotted as a function of $h^2 \OGW(f)$ and $\mathcal{A}_\textrm{lin}$. The remaining model parameters are as in \eqref{eq:sharp-post-benchmark-II}. The black contours denote lines of constant SNR.}
\label{fig:sharp_post_lognorm_Alin_logObar_3_0p1_21_pio2}
\end{figure}

We can extend the above analysis by scanning over values of $h^2 \ObarGW$ and $\mathcal{A}_\textrm{lin}$, keeping the remaining parameters as in \eqref{eq:sharp-post-benchmark-I}, and compute the SNR and the relative error $\sigma(\mathcal{A}_\textrm{lin}) / \mathcal{A}_\textrm{lin}$ for every combination. For the sharp feature template with flat envelope considered here the results are shown in fig.~\ref{fig:sharp_post_flat_Alin_logObar_21_pio2}. The shaded areas denote different decades of the relative error in $\%$ and the black contours are lines of constant SNR. As expected, the SNR is effectively independent of $\mathcal{A}_\textrm{lin}$, and strongly depends on the overall amplitude $h^2 \ObarGW$. Again, for $h^2 \ObarGW \lesssim 10^{-13}$ we expect that the signal may not be detectable as the SNR becomes too small. From the plot we can read off the minimal value of $h^2 \ObarGW$ needed to determine $\mathcal{A}_\textrm{lin}$ with a $\%$-accuracy. E.g.~if we wish to determine $\mathcal{A}_\textrm{lin}$ with better than 10$\%$ accuracy, for $\mathcal{A}_\textrm{lin}=0.4$ we require $h^2 \ObarGW > 10^{-12.3}$, while for $\mathcal{A}_\textrm{lin}=0.05$ we require $h^2 \ObarGW > 10^{-11.3}$, i.e.~the overall amplitude needs to be enhanced by one decade. Overall, one finds that when $\mathcal{A}_\textrm{lin}$ is decreased from $\mathcal{A}_\textrm{lin} = 0.2$ the required overall amplitude $h^2 \ObarGW$ for a given accuracy in $\mathcal{A}_\textrm{lin}$ quickly grows. Still, there is a sizable region in fig.~\ref{fig:sharp_post_flat_Alin_logObar_21_pio2} where the expected uncertainty in $\mathcal{A}_\textrm{lin}$ is $< 1 \%$.  

\begin{figure}[t]
\centering
\begin{overpic}[width=0.98\textwidth]{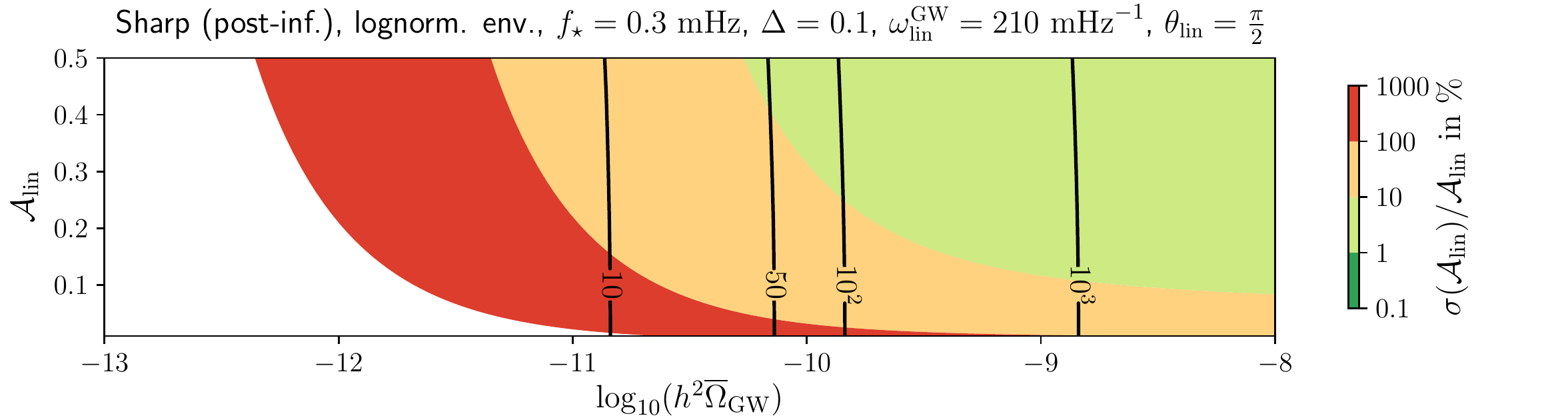}
\end{overpic}
\caption{As fig.~\ref{fig:sharp_post_lognorm_Alin_logObar_3_0p1_21_pio2}, but for model parameters as in \eqref{eq:sharp-post-benchmark-III}.}
\label{fig:sharp_post_lognorm_Alin_logObar_0p3_0p1_210_pio2}
\end{figure}

\begin{figure}[t]
\centering
\begin{overpic}[width=0.98\textwidth]{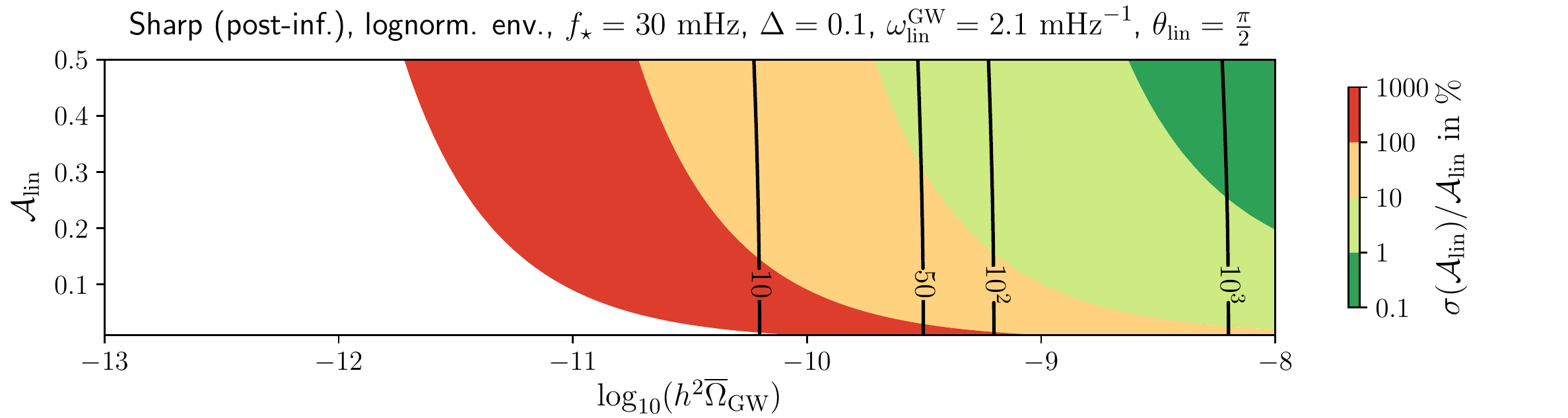}
\end{overpic}
\caption{As fig.~\ref{fig:sharp_post_lognorm_Alin_logObar_3_0p1_21_pio2}, but for model parameters as in \eqref{eq:sharp-post-benchmark-IV}.}
\label{fig:sharp_post_lognorm_Alin_logObar_30_0p1_2p1_pio2}
\end{figure}

\subsubsection{Lognormal envelope}
\label{sec:fisher-sharp-post-lognormal}
Note that the toy example with a flat envelope considered above represents an optimistic case for the detectability of sharp-feature-type oscillations, when compared to a peaked template with the same overall amplitude at the maximum. However, it is peaked envelopes that one expects in practice for inflation models with sharp features and we hence now turn to this case. The sensitivity of LISA in this case will also depend on the locus in $f$-space of this peak, i.e.~whether it falls into the frequency range of maximal sensitivity or not. Here, we begin with the most optimistic case, and consider models with a peak in $h^2 \OGW(f)$ at $f_\star=3 \textrm{ mHz}$, i.e.~in the region of maximal LISA sensitivity. We hence consider the template \eqref{eq:GW-sharp-template}, but now with a lognormal envelope \eqref{eq:lognorm-env} and again perform a scan over $h^2 \ObarGW$ and $\mathcal{A}_\textrm{lin}$ and record the relative error in $\mathcal{A}_\textrm{lin}$ and the SNR. For the remaining model parameters we choose:
\begin{align}
\label{eq:sharp-post-benchmark-II} f_\star = 3 \textrm{ mHz} \, , \quad \Delta=0.1 \, , \quad \omegagwlin = 21 \textrm{ mHz}^{-1}, \quad \theta_\textrm{lin}=\frac{\pi}{2} \, .
\end{align}
For a sharp feature, the width of the peak in $f$-space typically does not exceed the central value $f_\star$, which for the lognormal template implies $\Delta < 1$. Here, to be specific, we choose $\Delta =0.1$, corresponding to a fairly narrow peak, also to highlight the difference to the flat envelope case studied before. The value $\omegagwlin = 21 \textrm{ mHz}^{-1}$ is as for the flat-envelope example, but now we can check that for this choice we have $\omegagwlin f_\star \Delta \approx 2 \pi$. As explained at the end of section \ref{sec:templates-sharp-post}, this is close to the minimal value for the template \eqref{eq:GW-sharp-template} to be valid, and allows for the largest range of possible $\mathcal{A}_\textrm{lin}$.\footnote{For $\omegagwlin f_\star \Delta \gg 2 \pi$ one would in general only expect small values of $\mathcal{A}_\textrm{lin} < \mathcal{O}(10 \%)$, and hence one would have to consider a more restricted range of values of $\mathcal{A}_\textrm{lin}$ to scan over.} An example GW spectrum with these parameters and $h^2 \ObarGW = 10^{-10.6}$ and $\mathcal{A}_\textrm{lin}=0.2$ is shown in fig.~\ref{fig:sharp_post_examples} (red curve).

In fig.~\ref{fig:sharp_post_lognorm_Alin_logObar_3_0p1_21_pio2} we plot the $\%$-error in $\mathcal{A}_\textrm{lin}$ and the SNR as a function of $h^2 \ObarGW$ and $\mathcal{A}_\textrm{lin}$ for this example with a lognormal peak with the remaining parameters given in \eqref{eq:sharp-post-benchmark-II}. It is most instructive to compare this to the corresponding plot in fig.~\ref{fig:sharp_post_flat_Alin_logObar_21_pio2} for a flat envelope. As expected, for the peaked example the SNR is generally lower and the $\%$-error larger for the same values of $h^2 \ObarGW$ and $\mathcal{A}_\textrm{lin}$. For example, to again have an error of $< 10\%$ and $\mathcal{A}_\textrm{lin}=0.4$ one requires $h^2 \ObarGW > 10^{-11.9}$, while for $\mathcal{A}_\textrm{lin}=0.05$ one requires $h^2 \ObarGW > 10^{-10.6}$. Remarkably, this is only a slight deterioration of the accessible parameter space compared to the flat-envelope case. However, the region where $\mathcal{A}_\textrm{lin}$ is expected to be determined with accuracy $<1\%$ is now inaccessible for $\mathcal{A}_\textrm{lin}< 0.2$ for the whole range of $h^2 \ObarGW$ plotted. 

We now consider how this situation changes if the peak does not fall into the `sweet spot' of maximal LISA sensitivity. For example, we now assume that the central value of the peak appears at a frequency lower or higher by a factor of 10 compared to the optimal case just studied. We choose parameters such that for shared values of $h^2 \ObarGW$ and $\mathcal{A}_\textrm{lin}$ we have $h^2 \OGW^{\eqref{eq:sharp-post-benchmark-II}}(f) = h^2 \OGW^{\eqref{eq:sharp-post-benchmark-III}}(f/10) = h^2 \OGW^{\eqref{eq:sharp-post-benchmark-IV}}(10f)$, i.e.:
\begin{align}
\label{eq:sharp-post-benchmark-III} & f_\star = 0.3 \textrm{ mHz} \, , \quad \Delta=0.1 \, , \quad \omegagwlin = 210 \textrm{ mHz}^{-1}, \quad \theta_\textrm{lin}=\frac{\pi}{2} \, , \\
\label{eq:sharp-post-benchmark-IV} & f_\star = 30 \textrm{ mHz} \, , \ \quad \Delta=0.1 \, , \quad \omegagwlin = 2.1 \textrm{ mHz}^{-1}, \quad \, \theta_\textrm{lin}=\frac{\pi}{2} \, .
\end{align}
Note that these choices maintain the relation $\omegagwlin f_\star \Delta \approx 2 \pi$. For illustration, in fig.~\ref{fig:sharp_post_examples} we plot two GW spectra for the parameter choices \eqref{eq:sharp-post-benchmark-III} and \eqref{eq:sharp-post-benchmark-IV} with $h^2 \ObarGW = 10^{-10.6}$ and $\mathcal{A}_\textrm{lin}=0.2$ (brown and violet curves). 

The corresponding plots of the $\%$-error in $\mathcal{A}_\textrm{lin}$ and the SNR are shown in figs.~\ref{fig:sharp_post_lognorm_Alin_logObar_0p3_0p1_210_pio2} and \ref{fig:sharp_post_lognorm_Alin_logObar_30_0p1_2p1_pio2}. These show a considerable deterioration of the prospects of detecting the signal and of measuring the sharp-feature-type oscillations compared to the case where the peak is centred at $f_\star = 3 \textrm{ mHz}$. E.g., for a $\textrm{SNR} >10$ one now requires $h^2 \ObarGW > 10^{-10.9}$ or $h^2 \ObarGW > 10^{-10.2}$ for the parameters \eqref{eq:sharp-post-benchmark-III} and \eqref{eq:sharp-post-benchmark-IV}, respectively. Similarly, for detecting $\mathcal{A}_\textrm{lin}=0.4$ with accuracy $< 10 \%$ one now requires $h^2 \ObarGW > 10^{-10.1}$ or $h^2 \ObarGW > 10^{-9.6}$. There are also interesting differences between the two plots. For a given value of $h^2 \ObarGW$ the SNR in fig.~\ref{fig:sharp_post_lognorm_Alin_logObar_30_0p1_2p1_pio2} is generally lower than in fig.~\ref{fig:sharp_post_lognorm_Alin_logObar_0p3_0p1_210_pio2}. In contrast, in fig.~\ref{fig:sharp_post_lognorm_Alin_logObar_30_0p1_2p1_pio2} there is a region in parameter space, where the uncertainty in $\mathcal{A}_\textrm{lin}$ is $< 1 \%$ while in fig.~\ref{fig:sharp_post_lognorm_Alin_logObar_0p3_0p1_210_pio2} such a region is absent. Similarly, from fig.~\ref{fig:sharp_post_lognorm_Alin_logObar_0p3_0p1_210_pio2} it follows that for the parameter choice \eqref{eq:sharp-post-benchmark-III} an oscillation with $\mathcal{A}_\textrm{lin}< 0.1$ cannot be determined with better than $\sim 10 \%$ accuracy for the whole range in $h^2 \ObarGW$ shown, while fig.~\ref{fig:sharp_post_lognorm_Alin_logObar_0p3_0p1_210_pio2} implies that for \eqref{eq:sharp-post-benchmark-IV} this is attainable for $h^2 \ObarGW \gtrsim 10^{-9}$.

The results of this section can be summarised as follows. According to the FIM analysis, a measurement of the oscillation in the SGWB induced by a sharp feature in the post-inflationary era by LISA is possible for a wide range of parameter values. Interestingly, the fact that the oscillating signal is typically peaked does not significantly reduce the detectability compared to a non-peaked signal, as long as the central value of the peak is close to the frequency range of maximal LISA sensitivity $f \sim 3 \textrm{ mHz}$. In this case an oscillating signal with oscillation amplitude $\mathcal{A}_\textrm{lin} \sim \mathcal{O}(10 \%)$ is expected to be detectable with $< 10 \%$ accuracy as long as the overall amplitude is $h^2 \ObarGW \gtrsim 10^{-12}$-$10^{-11}$. The prospects for detection decrease significantly when the central value of the peak occurs away from the frequency range of max.~LISA sensitivity. Still, for oscillation amplitudes $\mathcal{A} > 10 \%$ there is still a non-negligible parameter space for detection. 

\subsection{Sharp feature: inflation-era contribution}
\label{sec:fisher-sharp-inf}
The template for the GW spectrum sourced  by a sharp feature during inflation is given in \eqref{eq:GW-inf-template} and depends on the three parameters $(h^2 \ObarGW, \, \etap, \, \omegalin)$.

A characteristic property of this template are $\mathcal{O}(1)$ oscillations with frequency $\omegalin$ along the UV tail of the spectrum, which distinguishes this from a template with just a peak in $h^2 \OGW(f)$. A detection of this oscillating tail would be a strong hint for a sharp feature active during inflation. However, the Fisher analysis presented here will not be sufficient to determine whether these oscillations will be detectable in LISA. The reason is that the parameter $\omegalin$ describing the oscillations also controls the location of the global maximum of the template \eqref{eq:GW-inf-template}, given by $f_\textrm{max} \simeq 6.223 / \omegalin$. If a Fisher forecast then reveals a small uncertainty in $\omegalin$, this does not necessarily imply that LISA will be able to measure the oscillations, but it is more likely that this is caused by the fact that LISA is sensitive to the global maximum of the signal. At the level of a Fisher analysis it is not obvious how to disentangle the detectability prospects of the maximum vs.~the oscillations, which motivates further more advanced investigations of this issue.

\begin{figure}[t]
\centering
\begin{subfigure}{.5\textwidth}
 \centering
   \begin{overpic}
[width=0.9\textwidth]{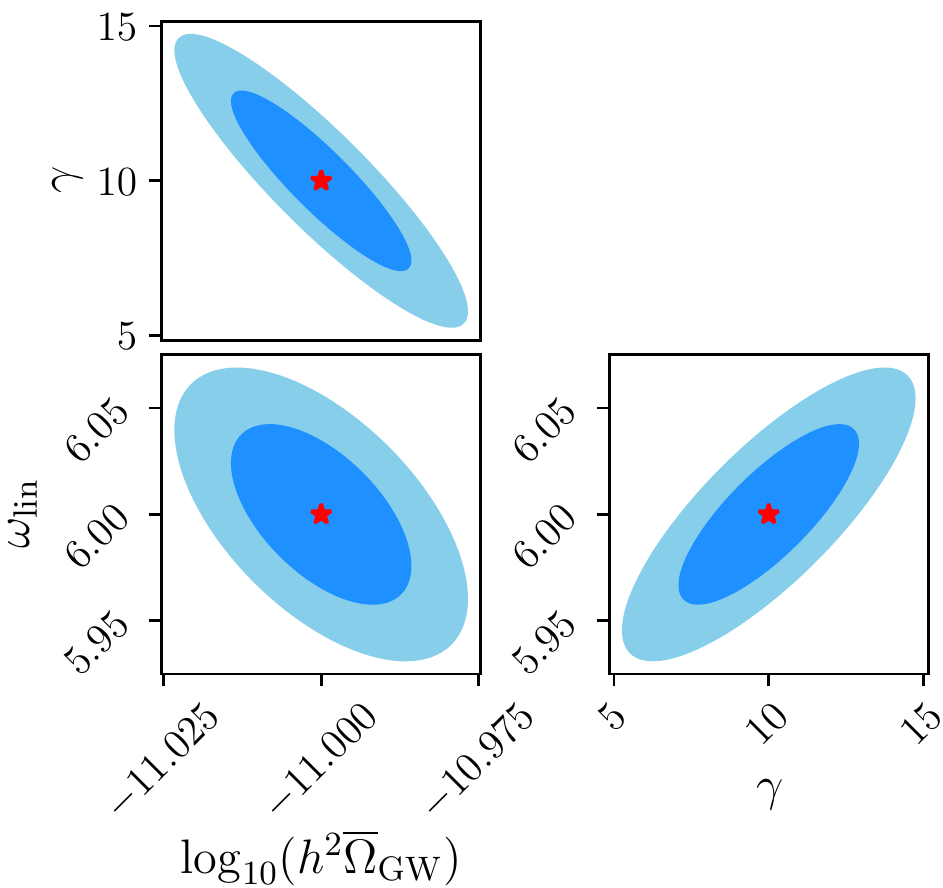}
\put(67,74){$\textrm{SNR}=178$}
\end{overpic}
\caption{\hphantom{A}}
\label{fig:corner_inf_m11_10_6}
\end{subfigure}%
\begin{subfigure}{.5\textwidth}
 \centering
   \begin{overpic}
[width=0.9\textwidth]{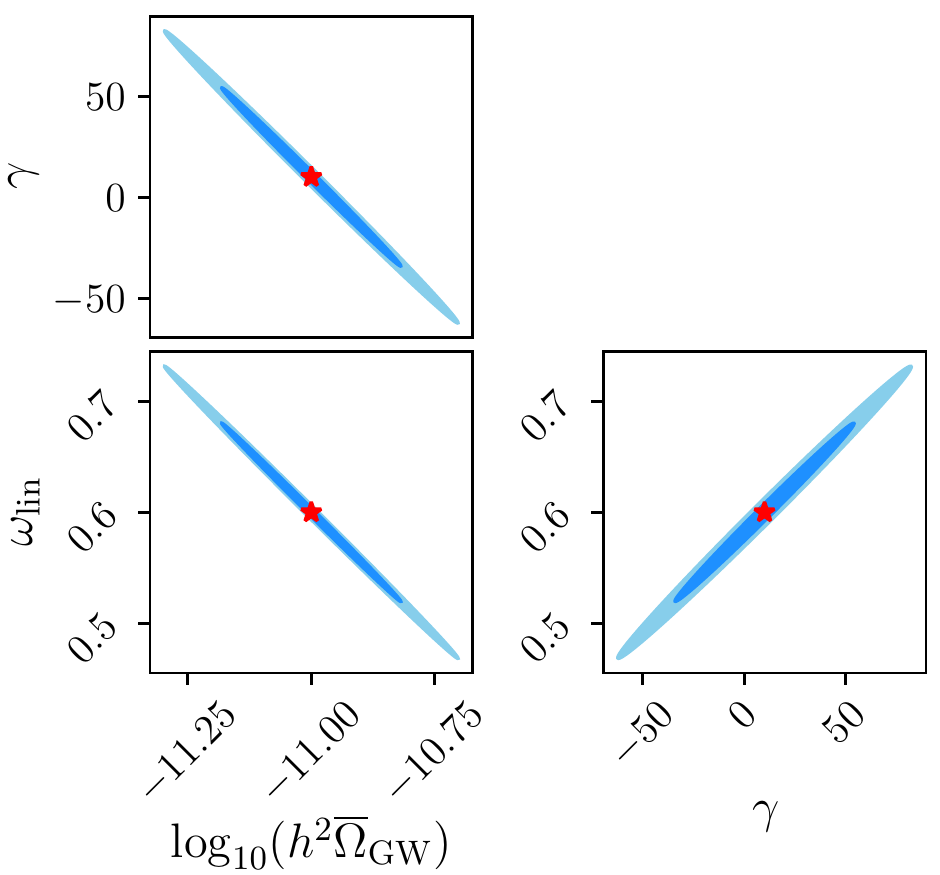}
\put(67,74){$\textrm{SNR}=137$}
\end{overpic}
\caption{\hphantom{A}}
\label{fig:corner_inf_m11_10_0p6}
\end{subfigure}%
\caption{Confidence ellipses with $1\sigma$ and $2\sigma$ regions obtained from a Fisher forecast for the inflation-era contribution to the SGWB due to a sharp feature with template \eqref{eq:GW-inf-template}. The model parameters are given in \eqref{eq:sharp-inf-benchmark-I} for panel \textbf{(a)} and \eqref{eq:sharp-inf-benchmark-II} for panel \textbf{(b)}.}
\label{fig:corner_inf}
\end{figure}

Still, one can reap the benefits of a Fisher forecast for the template \eqref{eq:GW-inf-template}. As this corresponds to a peaked signal, the detectability prospects will depend on the location in $f$-space of this peak, together with the maximal amplitude $h^2 \ObarGW$. As the peak location is controlled by $\omegalin$, one can stake out the accessible parameter space in $\omegalin$ as a function of $h^2 \ObarGW$. Any future analysis of the visibility of the oscillation can then be restricted to this most promising region of parameter space. In addition, we will use the Fisher analysis to make visible correlations between the parameters in \eqref{eq:GW-inf-template}.

We begin by plotting the confidence ellipses for the parameters for two example models with parameter choices:
\begin{align}
\label{eq:sharp-inf-benchmark-I}    & h^2 \ObarGW = 10^{-11} \, , \quad \etap=10 \, , \quad \omegalin= 6 \textrm{ mHz}^{-1} \, , \\
\label{eq:sharp-inf-benchmark-II}    & h^2 \ObarGW = 10^{-11} \, , \quad \etap=10 \, , \quad \omegalin= 0.6 \textrm{ mHz}^{-1} \, .
\end{align}
The corresponding GW spectra are shown in the left panel of fig.~\ref{fig:inf_sharp_res_post_examples}. The choice $\gamma=10$ is consistent with the requirement of validity of the expression \eqref{eq:GW-inf-template}, but we will see that the exact value will not be very important as this parameter will be difficult to measure precisely. The choice $\omegalin= 6 \textrm{ mHz}^{-1}$ in \eqref{eq:sharp-inf-benchmark-I} was made so that the global maximum of the GW spectrum occurs at $f_\textrm{max} \simeq 1 \textrm{ mHz}$ and hence close to the region of max.~LISA sensitivity. The value in \eqref{eq:sharp-inf-benchmark-II} was chosen so that $f_\textrm{max}$ is somewhat displaced from this region towards larger values of $f$. 

The confidence ellipses for the two parameter choices in \eqref{eq:sharp-inf-benchmark-I} and \eqref{eq:sharp-inf-benchmark-II} are shown in figs.~\ref{fig:corner_inf_m11_10_6} and \ref{fig:corner_inf_m11_10_0p6}, respectively. We make the following observations. Unsurprisingly, the uncertainties in the parameters are smaller for the example spectrum that peaks close to the region of max.~LISA sensitivity. Also, out of the three parameters, the largest relative error generically occurs for $\etap$. This is expected as $\gamma$ only controls the shape of the lower UV tail of the GW spectrum, which is difficult to access experimentally. 

\begin{figure}[t]
\centering
\begin{overpic}[width=0.98\textwidth]{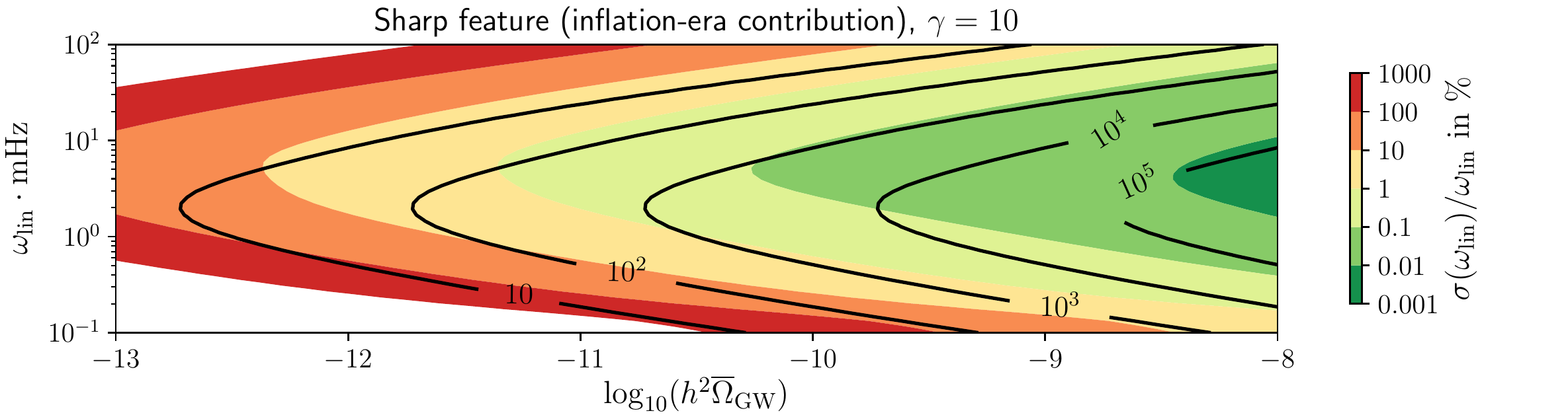}
\end{overpic}
\caption{Relative error $\sigma(\omegalin) / \omegalin$ in $\%$ from a Fisher forecast for the inflationary era GW spectrum due to a sharp feature with template \eqref{eq:GW-inf-template}, plotted as a function of $h^2 \OGW(f)$ and $\omegalin$ for $\gamma=10$. The black contours denote lines of constant SNR.}
\label{fig:sharp_inf_wlin_logObar_10}
\end{figure}

Interestingly, the Fisher analysis exhibits correlations between the three model parameters, which for the example in fig.~\ref{fig:corner_inf_m11_10_0p6} are severe. These can be understood by the fact that the template \eqref{eq:GW-inf-template} is not symmetric about its maximum (in $\ln(f)$-space), but has more `power' on the IR side than in the UV tail which is modulated by the oscillations, see e.g.~the left panel of fig.~\ref{fig:inf_sharp_res_post_examples}. 
For the example in fig.~\ref{fig:corner_inf_m11_10_0p6}, LISA is mainly sensitive to the IR slope of the signal, as the peak is shifted to the right compared to the region of max.~LISA sensitivity. To get the dependence of the IR tail on the model parameters, we expand the template for small $\omegalin f$, finding at leading order $h^2 \OGW(f) \sim h^2 \ObarGW \, (\omegalin f)^3$. Thus the IR slope is controlled by the combination $h^2 \ObarGW \, \omegalin^3$, which explains the strong anti-correlation between $h^2 \ObarGW$ and $\omegalin$ in fig.~\ref{fig:corner_inf_m11_10_0p6}. Then, note that $\etap$ only appears in the template \eqref{eq:GW-inf-template} in the combination $\omegalin / \etap$, which is the cause for the correlation between these two parameters. The correlations between the pairs $(h^2 \ObarGW, \, \omegalin)$ and $(\omegalin, \, \etap)$ then imply a correlation between $(h^2 \ObarGW, \, \etap)$. For the example in fig.~\ref{fig:corner_inf_m11_10_6} these correlations are still visible, but they are weaker, as the IR tails plays a less important role for the detectability of the signal when the peak appears centrally in the region of max.~LISA sensitivity. Still, the importance of the IR slope of the signal compared to the UV tail implies that these correlations are not completely absent.\footnote{One can check that when the peak in $h^2 \OGW(f)$ is shifted to smaller values $f < 1 \textrm{ mHz}$, the anti-correlation between $h^2 \ObarGW$ and $\omegalin$ eventually becomes a correlation, as the IR tails becomes `unobservable', and the more important part becomes the UV tail, which just beyond the maximum scales as $h^2 \OGW(f) \sim h^2 \ObarGW \, (\omegalin f)^{-3}$. However, we also find that when the peak is shifted to smaller values of $f$ the SNR drops very fast, making this region of parameter space unpromising for detection.} 

In fig.~\ref{fig:sharp_inf_wlin_logObar_10} we then plot the relative error in $\omegalin$ and contours of constant SNR as a function of $h^2 \ObarGW$ and $\omegalin$. The shapes of the contours follow from the fact that for a given $h^2 \ObarGW$ both the SNR is maximized as well as the error in $\omegalin$ minimized when the peak occurs in the region of max.~LISA sensitivity. In particular, for fixed $h^2 \ObarGW$ the SNR is largest for $\omegalin \simeq 2 \textrm{ mHz}^{-1}$, which corresponds to the global maximum of the GW spectrum lying on the LISA `sweet spot' at $f_\textrm{max} \simeq 3 \textrm{ mHz}$. The fact that the minimum $\%$-error occurs at slightly larger values of $\omegalin$ is caused by the fact that the relative error goes down as $\omegalin$ is increased, as long as the absolute error does not change much. According to fig.~\ref{fig:sharp_inf_wlin_logObar_10}, for $h^2 \ObarGW = 10^{-12}$ a better-than-$10 \%$ measurement of $\omegalin$ is possible for $\omegalin$ in the range $\omegalin \in [1.7, \, 12.5] \textrm{ mHz}^{-1}$, while for $h^2 \ObarGW = 10^{-10}$ this band expands to $\omegalin \in [0.26, \, 78] \textrm{ mHz}^{-1}$. In a follow-up analysis dedicated to studying the observability of the oscillations on the UV tail one can then focus on these parameter ranges most relevant for detection.

\begin{figure}[t]
\centering
\begin{overpic}[width=0.98\textwidth]{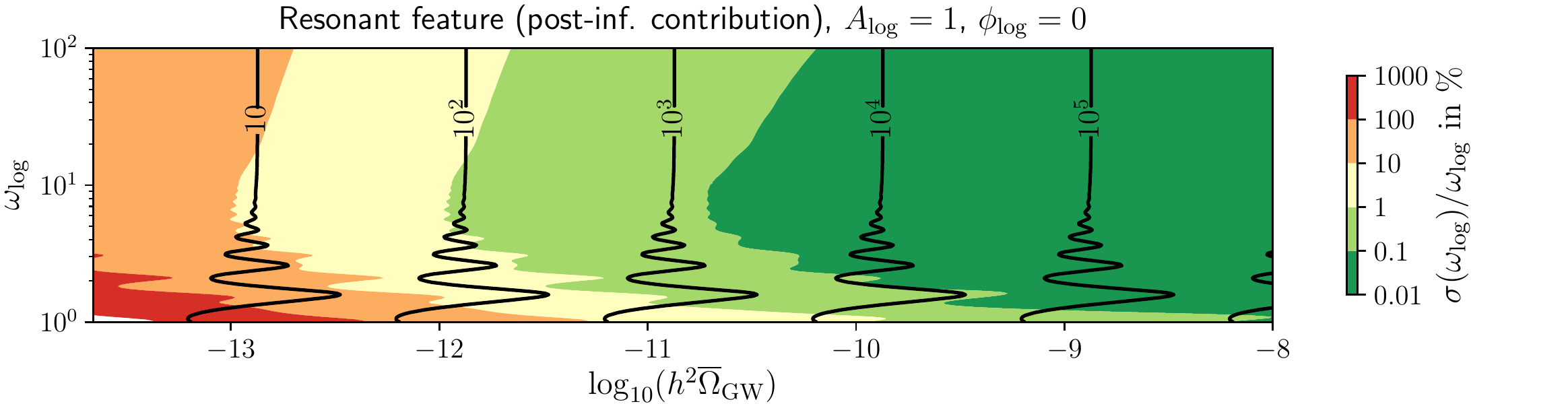}
\end{overpic}
\caption{Relative error $\sigma(\omegalog) / \omegalog$ in $\%$ from a Fisher forecast for a resonant feature with template \eqref{eq:resonant-template-A}, plotted as a function of $h^2 \OGW(f)$ and $\mathcal{A}_\textrm{lin}$. The remaining model parameters are $A_\textrm{log}=1$ and $\phi_\textrm{log}=0$. The black contours denote lines of constant SNR.}
\label{fig:sharp_post_res_wlog_logObar_1_0}
\end{figure}

\begin{figure}[t]
\centering
\begin{overpic}[width=0.98\textwidth]{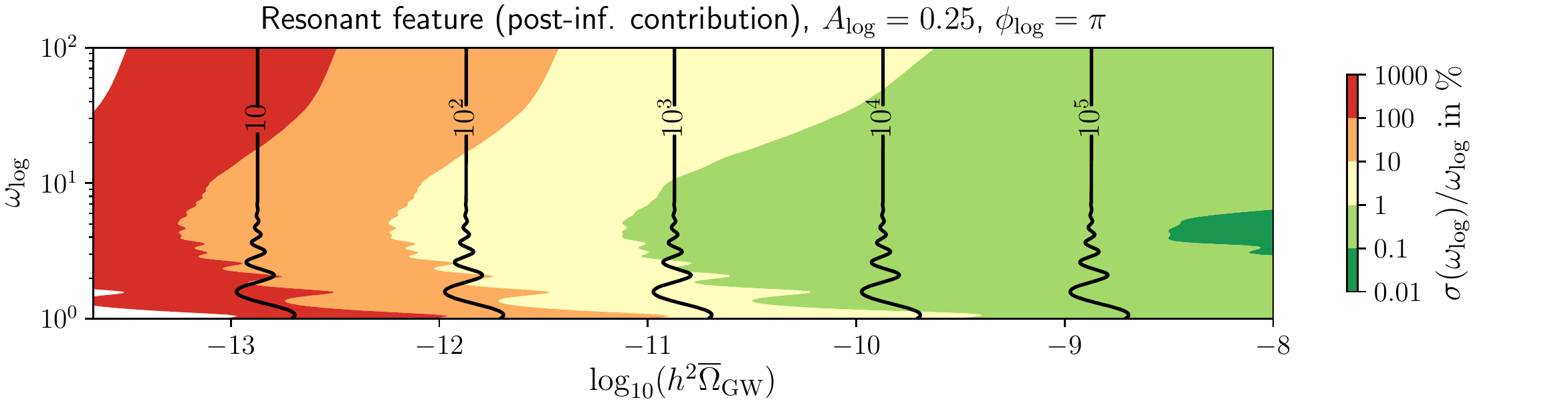}
\end{overpic}
\caption{As fig.~\ref{fig:sharp_post_res_wlog_logObar_1_0}, but for $A_\textrm{log}=0.25$ and $\phi_\textrm{log}=\pi$.}
\label{fig:sharp_post_res_wlog_logObar_0p25_pi}
\end{figure}

\subsection{Resonant feature: post-inflationary contribution}
\label{sec:fisher-resonant-post}
We now turn to the post-inflationary contribution to the SGWB due to a resonant feature during inflation with template \eqref{eq:resonant-template-A}. Recall that this template is valid as long as the envelope of $\Pzeta$ for the scales affected by the resonant feature is approximately constant, at least over several periods of the oscillation. 

Once more, our main interest is to determine to what extent the oscillations characteristic for this type of feature can be detected with LISA. To this end we will focus on the predicted accuracy for measuring $\omegalog$, as this parameter controls the main properties of the oscillation in the template \eqref{eq:resonant-template-A}: it sets both the frequency of the oscillating pieces and also their relative amplitudes through the functions $\C_{0,1,2}(\omegalog)$.\footnote{The phase $\phi_\textrm{log}$ is less important, as it only shifts the signal in $f$-space, but it will play a role for the detectability for GW signals with small values of $\omegalog$, where $\phi_\textrm{log}$ will determine whether a maximum or a minimum of the oscillations will occur in the frequency region of max.~LISA sensitivity.} In fig.~\ref{fig:sharp_post_res_wlog_logObar_1_0} we show the results from a Fisher forecast for the predicted relative error in $\omegalog$ as a function of $\omegalog$ and the overall amplitude $h^2 \ObarGW$. For the other model parameters we choose  $A_\textrm{log}=1$ and $\phi_\textrm{log}=0$. Recall that $A_\textrm{log}$ sets the amplitude of the oscillations in the \emph{scalar} power spectrum, and here, for a first analysis, we have chosen the most optimistic value for detection. 

We make the following observations. One conspicuous property are the oscillations in the required value of $h^2 \ObarGW$ for a given $\%$-error or a given SNR for $\omegalog \lesssim 10$. This can be explained as follows. For $\omegalog \lesssim 10$ a period of the oscillations in the GW signal covers such an extended region in $f$-space, that at most a few and sometimes even fewer periods occur in the region of optimal LISA sensitivity. For example, for the oscillatory piece in \eqref{eq:resonant-template-A} with frequency $\omegalog$, a single period extends over more than a decade in $f$-space if $\omegalog < 2.73$. Then, if a maximum of the oscillation occurs in the region of max.~LISA sensitivity, this increases the overall signal in this frequency range, thus boosting the SNR, while the opposite occurs for minima.\footnote{As a result, the  wiggles in the SNR contours will also depend on the value of $\phi_\textrm{log}$, as the loci of maxima and minima of the oscillation will get shifted under a change of $\phi_\textrm{log}$.} For example, for the blue curve in the right panel of fig.~\ref{fig:inf_sharp_res_post_examples} a maximum of the oscillation coincides with the region of max.~LISA sensitivity, thus increasing the SNR. 

The oscillations in the contours of the $\%$-error are also in part caused by this effect. In addition, as $\omegalog$ is varied, the oscillatory pattern of $h^2 \OGW(f)$ changes as the relative amplitudes and phases of the two oscillatory pieces in \eqref{eq:resonant-template-A} shift with $\omegalog$. This affects the `visibility' of the frequency $\omegalog$ underlying the oscillations and hence the error fluctuates with $\omegalog$.         
For $\omegalog \gtrsim 8$ the SNR becomes effectively independent of $\omegalog$ and is determined by the value of $h^2 \ObarGW$, just like for the sharp feature template studied before in sec.~\ref{sec:fisher-sharp-post}. The reason is that for $\omegalog \gtrsim 8$ the LISA range in $f$-space covers many periods of the oscillation. In addition the oscillations are less pronounced, as the functions $\C_{1,2}(\omegalog)$ controlling the oscillation amplitude in \eqref{eq:resonant-template-A} decrease with growing $\omegalog$. Together, these make the `visibility' of the signal less dependent on the oscillations. The $\%$-error in $\omegalog$ reaches a minimum at $\omegalog \simeq 8$ (for fixed $h^2 \ObarGW$), and then grows as  $\omegalog$ is increased. This increase can be understood again by the decrease of $\C_{1,2}(\omegalog)$ with growing $\omegalog$. This suppresses the oscillation amplitude for large $\omegalog$, making an accurate measurement of $\omegalog$ increasingly difficult. An example of a resonant feature in $h^2 \OGW(f)$ with $\omegalog=8$ is shown as the red curve in the right panel of fig.~\ref{fig:inf_sharp_res_post_examples}.

For $A_\textrm{log}<1$ the error in $\omegalog$ is generally larger and the predicted accuracy also deteriorates faster with increasing $\omegalog$. This can be seen in fig.~\ref{fig:sharp_post_res_wlog_logObar_0p25_pi} where we plot the $\%$-error in $\omegalog$ as a function of $h^2 \ObarGW$ and $\omegalog$ for $A_\textrm{log}=0.25$ and $\philog=\pi$. The reason for this is that for $A_\textrm{log}<1$ the amplitude of oscillation is smaller, making the modulation harder to measure accurately. For large values of $\omegalog$ the oscillation in \eqref{eq:resonant-template-A} is in general dominated by the oscillatory piece with frequency $2 \omegalog$, as $\C_1(\omegalog)$ is suppressed compared to $\C_2(\omegalog)$ for $\omegalog >10$, see fig.~\ref{fig:C012theta12}. However, the oscillatory term with frequency $2 \omegalog$ term also comes with a prefactor $A_\textrm{log}^2$ that quickly suppresses the oscillation amplitude for $A_\textrm{log}<1$, and thus increases the error in $\omegalog$. In addition, the reduced amplitude of oscillation for $A_\textrm{log}=0.25$ leads to the wiggles in the SNR contours being less pronounced compared to what was observed for $A_\textrm{log}=1$. Also note that the wiggles in fig.~\ref{fig:sharp_post_res_wlog_logObar_0p25_pi} now start on the right from the centre when viewed from the bottom of the plot, whereas in fig.~\ref{fig:sharp_post_res_wlog_logObar_1_0} they started on the left. This is because what was a maximum of the oscillation across the range of max.~LISA sensitivity for $\philog=0$ has become a minimum for $\philog=\pi$ and vice versa. This shows the dependence of the SNR curve on $\philog$ for small $\omegalog$. Finally, the minimal error in $\omegalog$ for $A_\textrm{log}=0.25$ now occurs for $\omegalog \approx 5$. 

All in all, according to the Fisher forecast, an oscillation of resonant-feature-type as given by the template \eqref{eq:resonant-template-A} has the best prospects for detection with LISA for $\omegalog \approx 4$ to $10$, as the relative error in $\omegalog$ is generically smallest in this range independent of the other model parameters, see figs.~\ref{fig:sharp_post_res_wlog_logObar_1_0} and \ref{fig:sharp_post_res_wlog_logObar_0p25_pi}. For values of $\omegalog$ in this range, a better-than-1$\%$ accuracy requires $h^2 \ObarGW \gtrsim 10^{-12}$ for $A_\textrm{log}=1$ and $h^2 \ObarGW \gtrsim 10^{-11}$ for $A_\textrm{log}=0.25$. However, note that in the template \eqref{eq:resonant-template-A} the parameter $\omegalog$ sets both the frequency of the oscillation and it also controls the oscillation amplitude. As a consequence, the Fisher forecast does not provide conclusive results regarding the detectability of the signal: a small uncertainty in $\omegalog$ may signal that an oscillation with this frequency is in principal detectable, but it does not guarantee that the amplitude of oscillation can be resolved by the experiment.\footnote{Recall e.g.~the results from sec.~\ref{sec:fisher-sharp-post} for a sharp feature (post-inflationary contribution), where the error on $\omegagwlin$ is generically small, and the detectability of the signal appears instead to be limited by the accuracy in the measurement of $\mathcal{A}_\textrm{lin}$.} For conclusive results a more advanced analysis based on signal reconstruction would be needed. Initial results using the PCA method in sec.~\ref{sec:PCA} indicate that a reconstruction of resonant-feature-type oscillations with LISA is indeed possible. We take this as a strong motivation for further work in this direction. 

\section{Conclusions and outlook}
\label{sec:conclusions}
In this work we studied the detectability with LISA of contributions to the SGWB due to features during inflation, whose characteristic property are oscillations in the frequency profile. We distinguish between the post-inflationary era contribution to the SGWB where the oscillations modulate the peak of the GW spectrum \cite{Fumagalli:2020nvq, Fumagalli:2021cel, Witkowski:2021raz}, corresponding to the templates \eqref{eq:GW-sharp-template} and \eqref{eq:resonant-template-A}, and the inflationary-era contribution where the oscillations occur on the UV tail \cite{Fumagalli:2021mpc}, see the template \eqref{eq:GW-inf-template}.

Applying the PCA reconstruction technique of \cite{Pieroni:2020rob} to a selection of mock GW specta based on the templates \eqref{eq:GW-sharp-template}, \eqref{eq:GW-inf-template} and \eqref{eq:resonant-template-A} we find that the oscillations typical for features can in principle be reconstructed by LISA. Given this positive result, to go beyond individual examples and identify regions of parameter space that are promising for the detectability of the oscillations, we perform a Fisher forecast, while at the same time scanning over the parameters characterising the oscillation and the overall amplitude of the GW spectrum. Our results can be summarised as follows.
\begin{itemize}
\item For the \emph{post-inflationary era} contribution to $h^2 \OGW$ due to a \emph{sharp feature}, corresponding to the template \eqref{eq:GW-sharp-template}, an oscillating signal with oscillation amplitude $\mathcal{A}_\textrm{lin} \sim \mathcal{O}(10 \%)$ is expected to be measureable with $< 10 \%$ accuracy as long as the overall amplitude is $h^2 \OGW \gtrsim 10^{-12}$-$10^{-11}$, and the signal peaks close to the frequency range of maximal LISA sensitivity. For oscillation amplitudes $\mathcal{A}_\textrm{lin} < 10 \%$ and for signals that peak far away from the LISA `sweet spot' the detectability prospects quickly deteriorate. 
\item For the \emph{inflationary-era} contribution due to a \emph{sharp feature}, given by the template \eqref{eq:GW-inf-template}, the oscillation frequency is predicted to be measurable with $<10 \%$ accuracy in a band around $\omegalin \sim \mathcal{O}(1) \textrm{ mHz}^{-1}$. For these values the peak of the GW spectrum, which is also controlled  by $\omegalin$, lies within the LISA frequency range. The width of the $10\%$-accuracy band depends on the amplitude of $h^2 \OGW$ at the peak and roughly spans a factor $10$ in $\omegalin$ for $h^2 \OGW=10^{-12}$ and a factor $10^3$ for $h^2 \OGW=10^{-10}$. However, as the oscillations only occur on the UV tail of the GW spectrum, the Fisher analysis is not sufficient to determine whether the oscillations are detectable, as the uncertainty is largely determined by the visibility of the main peak, and more sophisticated reconstruction techniques are needed. Our initial results using the PCA method indicate that the oscillations are visible, as long as the overall amplitude of the GW signal is large enough, e.g.~for $h^2 \OGW \sim 10^{-10.5}$. 
\item For the \emph{post-inflationary era} contribution due to a \emph{resonant feature} given by the template \eqref{eq:resonant-template-A}, GW signals with oscillation frequency $\omegalog$ in the range $\omegalog \approx 4$ to $10$ have the best prospect for detection, with the lowest value of uncertainty for this parameter occurring in this band regardless of the other model parameters. In this $\omegalog$-band the two oscillatory pieces in the resonant feature template \eqref{eq:resonant-template-A} have a comparable amplitude, resulting in an intricate oscillation pattern. Signals with larger values of $\omegalog$ may still be detectable as long as the parameter $A_\textrm{log}$ controlling the oscillation amplitude in the associated scalar power spectrum \eqref{eq:P-resonant-intro} is close to unity. 
\end{itemize}

These results can then be compared to constraints coming from theoretical considerations of feature models. For scalar-induced GWs, the energy density fraction in GWs and the power spectrum of scalar fluctuations are related as $h^2 \OGW \sim 10^{-5} \Pzeta^2$. As a result, the largest attainable value of $h^2 \OGW$ is bounded by the maximal value of $\Pzeta$, which is constrained by considerations of backreaction and perturbative control. The exact values of these bounds will depend on the inflation model and in particular the mechanism responsible for enhancing the scalar fluctuations, see e.g.~\cite{Fumagalli:2020nvq, Inomata:2021zel, Inomata:2021tpx} for works where such issues are addressed. By comparing the required value $h^2 \OGW$ for detection with the maximal value from backreaction and perturbativity considerations one can identify which enhancement mechanisms are promising for detection via the SGWB, or will remain inaccessible to LISA.

The analysis in this work is only a first foray into the study of the detectability of oscillations in $h^2 \OGW(f)$, and there are many directions for improvement. For example, we assumed that the LISA noise parameters are known and did not include them in the FIM. We also neglected astrophysical foregrounds, which in the LISA frequency band are expected to be important and could dominate over the cosmological signal in some parts of the frequency band. 

Another limitation of the Fisher analysis is that it does not give information about the best-suited fiducial model to describe the dataset. To claim detection of a feature in GW data, the corresponding template with oscillations should be a better fit than another template, e.g.~a smooth template without oscillations. Thus, while the results of our Fisher analysis identify the most promising regions of parameter space for detecting features in the SGWB, they do not guarantee detection, especially for parameter values with a marginal uncertainty according to the Fisher forecast. 

Despite these inevitable limitations, the Fisher analysis in this work shows that oscillations in $h^2 \OGW$ due to features have good prospects of detection with LISA over large regions of parameter space, which is further supported by successfully reconstructed examples using the PCA technique. Overall, we take this as a strong motivation for further investigations regarding the detectability of these models, in particular using more sophisticated methods of signal reconstruction, like the binning method in \cite{Caprini:2019pxz, Flauger:2020qyi}. 

We conclude by mentioning that our analysis is specific to the LISA mission, and a similar study, which could also include cross-correlations~\cite{ Ruan:2020smc, Seto:2020zxw, Orlando:2020oko}, can be undertaken for the Taiji observatory \cite{Ruan:2018tsw}, which is projected to be also sensitive to the mHz range like LISA. The mHz range broadly corresponds to features during inflation that occur some $\sim 30$ $e$-folds after the CMB modes have exited the horizon, see e.g.~\cite{Fumagalli:2020nvq}. Yet, features can in principle occur from shortly after CMB mode exit to close to the end of inflation, and there is hence no preference for features to appear in the mHz range. Thus, it would be desirable to also perform corresponding detection forecasts for features in the SGWB for other GW observatories that cover other frequency ranges.  

\acknowledgments
We thank Gonzalo A. Palma, Spyros Sypsas and Cristobal Zenteno for the fruitful collaboration on a previous work that resulted in one of the templates used here. We are also very grateful to Matteo Braglia for useful comments on the draft of this paper. J.F, S.RP, and L.T.W are supported by the European Research Council under the European Union's Horizon 2020 research and innovation programme (grant agreement No 758792, project GEODESI). J.F is currently supported by a Contrato de Atracción de Talento (Modalidad 1) de la Comunidad de Madrid (Spain), number 2017-T1/TIC-5520 and the IFT Centro de Excelencia Severo Ochoa Grant SEV-2. The work of M.P was supported by STFC grants ST/P000762/1 and ST/T000791/1. M.P acknowledges support by the European Union's Horizon 2020 Research Council grant 724659 MassiveCosmo ERC- 2016-COG.

\appendix

\section{Resonant feature: numerical coefficients}
\label{app:resonant-template-coeffs}
In this appendix we report how the numerical coefficients $\C_{0,1,2}(\omegalog)$ and $\theta_{\textrm{log}, 1,2} (\omegalog)$ in \eqref{eq:resonant-template-A} are computed. Following \cite{Fumagalli:2021cel} they can be written as:
\begin{align}
\nonumber
&\C_0 (\omegalog) \equiv \frac{C_2 (\omegalog)}{A_0} , \quad \C_1 \equiv \frac{\big(A_1(\omegalog)^2+B_1(\omegalog)^2\big)^{1/2}}{A_0} , \quad \C_2 \equiv \frac{\big(A_2(\omegalog)^2+B_2(\omegalog)^2\big)^{1/2}}{A_0} , \\
\label{eq:C012_tan1_and_tan2_def}
&\tan \big( \theta_1(\omegalog) \big)= \frac{B_1(\omegalog)}{A_1(\omegalog)} - \omegalog \ln 2 \, , \qquad \tan \big( \theta_2(\omegalog) \big)= \frac{B_2(\omegalog)}{A_2(\omegalog)} - \omegalog \ln 2 \, ,
\end{align}
where the functions $A_{0,1,2}$, $B_{1,2}$ and $C_2$ are given byu:
\begin{align}\label{appeq:A0}
    A_0 &= \int_0^{1} \textrm{d} d \int_{1}^\infty \textrm{d} s \, \mathcal{T} (d,s) ,
\end{align}
\begin{align}\label{appeq:A1}
    A_1(\omegalog) &= \int_0^{1} \textrm{d} d \int_{1}^\infty \textrm{d} s \, \mathcal{T} (d,s) \, \Big[ \cos \big( \omegalog \log(s+d) \big) + \cos \big( \omegalog \log(s-d) \big) \Big],\\
    \label{eq:B1}
    B_1(\omegalog) &= \int_0^{1} \textrm{d} d \int_{1}^\infty \textrm{d} s \, \mathcal{T} (d,s) \, \Big[\sin \big( \omegalog \log(s+d) \big) + \sin \big( \omegalog \log(s-d) \big) \Big],
\end{align}
\begin{align}\label{appeq:A2}
    A_2(\omegalog) &= \frac{1}{2} \int_0^{1} \textrm{d} d \int_{1}^\infty \textrm{d} s \, \mathcal{T} (d,s) \, \cos\big(\omegalog \log (s^2-d^2)\big),\\
    \label{appeq:B2}
    B_2(\omegalog) &= \frac{1}{2} \int_0^{1} \textrm{d} d \int_{1}^\infty \textrm{d} s \, \mathcal{T} (d,s) \, \sin\big(\omegalog \log (s^2-d^2)\big),\\
    \label{appeq:C2}
    C_2(\omegalog) &= \frac{1}{2} \int_0^{1} \textrm{d} d \int_{1}^\infty \textrm{d} s \, \mathcal{T} (d,s) \, \cos\bigg(\omegalog \log \frac{s+d}{s-d}\bigg) .
\end{align}
For the integration kernel $\mathcal{T}(d,s)$ one has to use the appropriate expression that appears in the formula for computing $\OGW$, i.e.:
\begin{align} \label{appeq:OmegaGW-std-expr}
\Omega_{\textrm{GW}}(f)&= \int_1^{\infty} \textrm{d} s \int_{0}^{1} \textrm{d} d \, \mathcal{T}(d,s) \, \mathcal{P}_{\zeta} \bigg(\frac{f}{2}(s+d)\bigg) \mathcal{P}_{\zeta} \bigg(\frac{f}{2}(s-d)\bigg) \, .
\end{align}
The kernel depends on the equation of state $w$ of the universe and the propagation speed $c_s$ of scalar fluctuations during the period that the post-inflationary GWs are induced. For the case of radiation domination ($w=c_s^2=1/3$), as we assume in this work, the expression for $\mathcal{T}(d,s)$ has been computed in \cite{Espinosa:2018eve,Kohri:2018awv}. The corresponding expressions for general values $w < 0 < 1$ and $c_s \leq 1$ are given in \cite{Domenech:2019quo,Domenech:2020kqm,Domenech:2021ztg}.\footnote{For $w \neq 1/3$ the expression for $\OGW(f)$ in \eqref{appeq:OmegaGW-std-expr} has to be multiplied by a factor $(f / f_\textrm{rh})^{-2b}$ with $b=(1-3w)/(1+3w)$, see e.g.~\cite{Witkowski:2021raz}, where $f_\textrm{rh}$ is the frequency that corresponds to the time of reheating, which in this case is assumed to be effectively instantaneous. For $w \neq 1/3$ this factor should be included in the template \eqref{eq:resonant-template-A}.}

\bibliographystyle{apsrev4-1}
\bibliography{Biblio-2021}
\end{document}